\documentclass[format=acmsmall,screen=true]{acmart}
\pdfoutput=1

\setcopyright{none}
\AtBeginDocument{%
  \providecommand\BibTeX{{%
    \normalfont B\kern-0.5em{\scshape i\kern-0.25em b}\kern-0.8em\TeX}}}

\acmDOI{}
\acmYear{2020}
\acmMonth{8}

\usepackage{amssymb, amsthm, amsfonts, amscd}

\usepackage{stmaryrd}
\usepackage{color}
\usepackage{bm}
\usepackage{listings}
\usepackage{booktabs}
\usepackage{hyperref}

\usepackage{subcaption}

\usepackage{bussproofs,varwidth}
\newenvironment{mathprooftree}
  {\varwidth{.9\textwidth}\centering\leavevmode}
  {\DisplayProof\endvarwidth}

\usepackage{framed}

\definecolor{strings}{rgb}{.624,.251,.259}
\definecolor{keywords}{rgb}{.224,.451,.686}
\definecolor{comment}{rgb}{.5,.5,.5}
\definecolor{meta}{rgb}{.581,.067,.0}
\definecolor{copyedit}{rgb}{1.0,.0,.0}

\def\R{\mathbb{R}}

\def\disj{\vee}
\def\conj{\wedge}

\DeclareMathOperator*{\rewritesto}{%
    \text{\raisebox{-0.5ex}{%
        \rlap{$=\mathrel{\mkern-4mu}=$}%
        \raisebox{1ex}{\hspace{0pt}$\leadsto$}%
    }}%
}

\makeatletter
\newcommand{\oset}[3][0ex]{%
  \mathrel{\mathop{#3}\limits^{
    \vbox to#1{\kern-2\ex@
    \hbox{$\scriptstyle#2$}\vss}}}}
\makeatother
\newcommand{\harpoon}{\oset{\rightharpoonup}}

\makeatletter
\DeclareRobustCommand\bigop[1]{%
  \mathop{\vphantom{\sum}\mathpalette\bigop@{#1}}\slimits@
}
\newcommand{\bigop@}[2]{%
  \vcenter{%
    \sbox\z@{$#1\sum$}%
    \hbox{\resizebox{\ifx#1\displaystyle.9\fi\dimexpr\ht\z@+\dp\z@\relax}{!}{$\m@th#2$}}%
  }%
}
\makeatother

\newcommand{\gensymb}{\DOTSB\bigop{\boxplus}}

\DeclareMathOperator*{\gen}{\textcolor{keywords}{\gensymb}}
\DeclareMathOperator*{\exquant}{\textcolor{keywords}{\exists}}
\DeclareMathOperator*{\ssum}{\textcolor{keywords}{\sum}}

\newcommand{\ind}[1]{
  \color{keywords}\pmb{\left[\vphantom{#1}\right.}
    \normalcolor #1
  \color{keywords}\pmb{\left.\vphantom{#1}\right]}
    \normalcolor
}
\newcommand{\deriv}[2]{
  \color{meta}\mathbf{D}\left\llbracket
    \normalcolor #1
  \color{meta}\middle|
    \normalcolor #2
  \color{meta}\right\rrbracket
    \normalcolor
}
\newcommand{\derivT}[2]{
  \color{meta}\mathbf{D^T}\left\llbracket
    \normalcolor #1
  \color{meta}\middle|
    \normalcolor #2
  \color{meta}\right\rrbracket
    \normalcolor
}
\newcommand{\inner}[2]{
  \color{meta}\left\langle
    \normalcolor #1
  \color{meta}\middle|
    \normalcolor #2
  \color{meta}\right\rangle
    \normalcolor
}
\newcommand{\EvalExpr}[2]{
  \color{meta}\left\llbracket
    \normalcolor #1
  \color{meta}\middle|
    \normalcolor #2
  \color{meta}\right\rrbracket
    \normalcolor
}
\newcommand{\CostExpr}[2]{
  \color{meta}\$\!\left\llbracket
    \normalcolor #1
  \color{meta}\middle|
    \normalcolor #2
  \color{meta}\right\rrbracket
    \normalcolor
}
\newcommand{\SpanCost}[2]{
  \color{meta}\widehat{\$}\!\left\llbracket
    \normalcolor #1
  \color{meta}\middle|
    \normalcolor #2
  \color{meta}\right\rrbracket
    \normalcolor
}
\newcommand{\SSABracket}[1]{
  \color{meta}\left\llbracket
    \normalcolor #1
  \color{meta}\right\rrbracket_{SSA}
    \normalcolor
}
\DeclareMathOperator*{\msum}{\textcolor{meta}{\sum}}
\DeclareMathOperator*{\mplus}{\textcolor{meta}{+}}
\DeclareMathOperator*{\mzero}{\textcolor{meta}{0}}
\newcommand{\msub}[1]{\textcolor{meta}{[} #1 \textcolor{meta}{]}}
\newcommand{\kwd}[1]{\operatorname{\textcolor{keywords}{\mathbf{#1}}}}
\newcommand{\keq}{\operatorname{\textcolor{keywords}{\texttt{=}}}}
\newcommand{\mwd}[1]{\operatorname{\textcolor{meta}{\mathbf{#1}}}}
\newcommand{\meq}{\operatorname{\textcolor{meta}{\texttt{=}}}}

\newcommand{\pluseq}{\mathrel{+}=}

\numberwithin{equation}{section}
\theoremstyle{plain} 
\newtheorem{thm}[equation]{Theorem}

\theoremstyle{definition} 
 
\theoremstyle{remark}

\lstdefinelanguage{lua}
  {morekeywords={and,break,do,else,elseif,end,false,for,function,if,in,local,
     nil,not,or,repeat,return,then,true,until,while,terra,var,let,quote,struct,
     view,set,field,seam,schema, into, operation, assert,
     emit, delete, new, invariant},
   morekeywords={[2]},
   morekeywords={[2]coroutine.create,coroutine.resume,coroutine.running,
     coroutine.status,coroutine.wrap,coroutine.yield},
   morekeywords={[2]module,require,package.cpath,package.load,package.loaded,
     package.loaders,package.loadlib,package.path,package.preload,
     package.seeall},
   morekeywords={[2]string.byte,string.char,string.dump,string.find,
     string.format,string.gmatch,string.gsub,string.len,string.lower,
     string.match,string.rep,string.reverse,string.sub,string.upper},
   morekeywords={[2]},
   morekeywords={[2]math.abs,math.acos,math.asin,math.atan,math.atan2,
     math.ceil,math.cos,math.cosh,math.deg,math.exp,math.floor,math.fmod,
     math.frexp,math.huge,math.ldexp,math.log,math.log10,math.max,math.min,
     math.modf,math.pi,math.pow,math.rad,math.random,math.randomseed,math.sin,
     math.sinh,math.sqrt,math.tan,math.tanh},
   morekeywords={[2]},
   morekeywords={[2]os.clock,os.date,os.difftime,os.execute,os.exit,os.getenv,
     os.remove,os.rename,os.setlocale,os.time,os.tmpname},
   sensitive=true,
   morecomment=[l]{--},
   morecomment=[s]{--[[}{]]},
   morestring=[b]",
   morestring=[d]',
   keywordstyle=\color{keywords},
   stringstyle=\color{strings},
   commentstyle=\color{comment}
  }

\usepackage{letltxmacro}
\newcommand*{\SavedLstInline}{}
\LetLtxMacro\SavedLstInline\lstinline
\DeclareRobustCommand*{\lstinline}{%
  \ifmmode
    \let\SavedBGroup\bgroup
    \def\bgroup{%
      \let\bgroup\SavedBGroup
      \hbox\bgroup
    }%
  \fi
  \SavedLstInline
}

\begin{document}

\title{Differentiating A Tensor Language}
\author{Gilbert Bernstein}
\email{gilbert@gilbertbernstein.com}
\affiliation{%
  \institution{University of California, Berkeley}
}

\author{Michael Mara}
\affiliation{%
  \institution{Stanford University}
}

\author{Tzu-Mao Li}
\affiliation{%
  \institution{Massachusetts Institute of Technology}
}

\author{Dougal Maclaurin}
\affiliation{%
  \institution{Google}
}

\author{Jonathan Ragan-Kelley}
\affiliation{%
  \institution{Massachusetts Institute of Technology}
}

\renewcommand{\shortauthors}{Bernstein, et al.}

\begin{abstract}
How does one compile derivatives of tensor programs, such that the resulting code is purely functional (hence easier to optimize and parallelize) and provably efficient relative to the original program?
We show that na\"ively differentiating tensor code---as done in popular systems like Tensorflow and PyTorch---can cause asymptotic slowdowns in pathological cases, violating the \emph{Cheap Gradients Principle}.
However, all existing automatic differentiation methods that guarantee this principle (for variable size data) do so by relying on $\pluseq$ mutation through aliases/pointers---which complicates downstream optimization.
We provide the first purely functional, provably efficient, adjoint/reverse-mode derivatives of array/tensor code by explicitly accounting for sparsity.  We do this by focusing on the indicator function from Iverson's APL.  We also introduce a new ``Tensor SSA'' normal form and a new derivation of reverse-mode automatic differentiation based on the universal property of inner-products.
\end{abstract}




\maketitle

\section{Introduction}

We present a means of statically computing provably efficient, purely functional derivatives of data-parallel code operating on both dense and sparse tensors (vectors, matrices, and other multi-dimensional array data).  This kind of code is important for a wide range of high-performance Domain-Specific Languages (DSLs) designed to solve optimization problems, including deep learning~\cite{Abadi:2015:TLM, Paszke:2019:PIS, Bergstra:2010:TCG}, non-linear least-squares~\cite{Devito:2017:ODS, ceres-solver}, inverse imaging problems~\cite{heide2016proximal, Li:2018:DPI}, and inverse simulation~\cite{hu2019difftaichi}.  All of these languages leverage automatic differentiation to shorten code, improve correctness and improve programmer productivity.  In all cases, the code produced by automatic differentiation sits in the inner loop of optimization algorithms, making overall performance highly dependent on the efficiency of the computer differentiated code and its compiler optimization---tasks which have largely been abstracted away from the client programmers.

Na\"ively differentiating tensor code can lead to asymptotically inefficient derivatives code, even with the help of reverse-mode automatic differentiation. Consider the following program written where we want to take the derivative of the output with respect to an input vector $x$ with size $N$: 
$$ \kwd{let} A \keq \text{diag}(x) \kwd{in} \text{trace}(A) + \cdots + \text{trace}(A). $$
Using common deep learning frameworks such as PyTorch or Tensorflow (without the XLA optimization), the original program runs in $O(N^2 + kN)$, where $k$ is the number of traces. In contrast, the gradient with respect to $A$ runs in $O(kN^2)$.  Thus, taking a ratio, the gradient is $O(k)$ times more expensive than the original function.  Why is this?  During normal evaluation, the trace operation reads and sums $n$ entries along the diagonal of a matrix.  However, when computing the gradient, the adjoint of trace is diag, which requires writing $n^2$ additional off-diagonal zeros to memory.
Standard algebraic simplification alleviates this problem (in this case a common subexpression elimination pass solves the problem), but no guarantee is provided.  For instance, turning on XLA\cite{xla} optimization will resolve this problem in Tensorflow.  However, now consider the following program:
$$ \kwd{let} A \keq \text{diag}(x) \kwd{in} \text{dot}(A[:,0], A[0,:])$$
where $A[:,0]$ means extract the first column of $A$ and $A[0,:]$ means extract the first row.  When compiled with XLA (via TensorFlow or JAX) this program runs in $O(N)$, but its gradient runs in $O(N^2)$.

Other works have also noted cases of asymptotic inefficiencies arising from algebraic simplifications (or lack thereof) in these systems~\cite{Shaikhha:2019:EDP, Laue:2018:CHO}.

Preserving parallelism in the differentiated tensor code is also non-trivial.
Prior automatic differentiation systems developed for scalar, imperative code (e.g.,~\cite{Utke:2008:OMO, Hascoet:2013:TAD}) do not produce asymptotically inefficient gradient code.
They follow the \emph{cheap gradient principle}~\cite{Griewank:2008:EDP}, which states that the time to compute the reverse-mode gradient of a function is at most a constant multiple of the time to compute the original function.
These systems achieve this by using the side-effecting $\pluseq$ reduction during reverse-mode execution.
Interestingly, most functional languages~\cite{Pearlmutter:2008:RAF,Li:2018:DPI,roesch2018relay,Wang:2019:DDP} also violate purity to include this key operation.  While we have no principled attachment to functional purity, in this case the introduction of side-effects destroys extant data-parallelism and obstructs downstream optimizations by indirecting dataflows through memory.  Significant and dramatic transformations of code are required to partially alleviate these issues, such as a scatter-to-gather loop optimization~\cite{Li:2018:DPI, Huckelheim:2019:ADA}.
To our knowledge, $\widetilde{F}$~\cite{Shaikhha:2019:EDP} is the only purely functional array language that does not rely on the $\pluseq$ reduction\footnote{Deep learning frameworks rely on imperative languages (e.g., CUDA) to implement the bulk operators and their derivatives.}.
Unfortunately, $\widetilde{F}$ lacks the crucial performance guarantee of cheap gradients.

We address the following question: ``\textbf{How do we compile the derivatives of a tensor program, such that the resulting derivative code is purely functional and provably efficient?}''. As just illustrated, the problem is that even when we restrict our attention to dense tensors, reverse-mode differentiation reveals \emph{sparse structures}.  In this work, we revisit an operator~\cite{Graham:1989:CMF} from Ken Iverson's APL~\cite{Iverson:1962:APL}: the indicator function $\ind{p}$ (Iverson's bracket) which evaluates to $1$ when $p$ is true and $0$ when $p$ is false.  We use this operator to link sparsity (in the form of boolean predicates) to linear algebra.  By carefully treating sparsity, we are able to show a memory-agnostic version of the cheap gradients principle: the number of new additions, multiplications, and non-linear operations introduced by the reverse-mode derivative is at most a constant multiple of the original function.  Importantly, we show that additions are only preserved \emph{non-locally} using a novel sparsity-aware proof.  

Our technical contributions are:
\begin{itemize}
    \item The first proof of the cheap gradient principle for a concise, purely functional array language that handles both dense and sparse tensors.
    \item A new, direct derivation of reverse-mode differentiation based on the universal-property of inner-products.
    \item An equivalent for SSA/A-normal form applicable to tensors; with which we derive worst case bounds for derivative transformations.
\end{itemize}

\textcolor{comment}{Note:  This manuscript is a draft of the formal arguments for this paper.  Please enjoy these preliminary findings.  We are working on experiments to buttress our claims, and will submit the entirety for peer review at that time.}

\section{Examples and Overview}
\label{sec:example-intro}

In this section, we will use some examples to illustrate our notation, how we derive reverse-mode (adjoint) derivatives, and the challenges of satisfying the cheap gradient principle in a purely functional setting.  These examples will mostly be derivatives of linear functions---which is not representative.  However, all of the important complexities (how to do reverse-mode, compose code, and handle sparsity/zeros) are well addressed by linear examples.

\subsection{A Convoluted Example}

We illustrate the notation of our language, and the derivation the derivatives using a 1D convolution program.
Consider the following tensor program that describes a 1D convolution (or stencil) $f: \R^{n+m} \rightarrow \R^{n}$:
\begin{equation*}
    f(x) = \gen_{i=0}^{n} \ssum_{j=0}^{m} x[i - j]\cdot c[j],
\end{equation*}
where $\gen_{i=0}^n e$ \emph{generates} an array $\left[e|_{i=0},e|_{i=1},\hdots,e|_{i=n-1}\right]$, and $\ssum_{j=0}^m e$ represents a summation over evaluations of the expression $e$ with $i$ varying $0$ to $m-1$. $x$ is the input signal and $c$ is a (constant) convolution kernel.

Let $\mathbf{D}f(x)$ denote the total derivative of $f$ (aka. the forward derivative) defined as the closest linear approximation to $f$ at base point $x$:
\begin{equation*}
    f(x + dx) \approx f(x) + \mathbf{D} f(x,dx) = f(x) + \mathbf{J}f(x) \cdot dx,
\end{equation*}
where $\mathbf{J} f(x)$ is the Jacobian of $f$.  Note the type, $\mathbf{D}f : \R^{n+m} \to (\R^{n+m} \to \R^n)$, where the nested function of $dx$ is necessarily linear. 

For our function $f(x) = e$, we use the notation $\mathbf{D}f(x,dx) = \deriv{e}{x\mapsto dx}$ to denote the meta-linguistic code transformation corresponding to forward differentiation.  By applying our rules (\S\ref{sec:derivatives}) we get the result:
\begin{equation*}
    \deriv{\gen_{i=0}^{n} \ssum_{j=0}^{m} x[i - j] c[j]}{x \mapsto dx}
        = \gen_{i=0}^{n} \ssum_{j=0}^{m} dx[i - j]\cdot c[j]
\end{equation*}

We can probe $\mathbf{D} f(x,dx)$ to see how infinitesimal changes to $x$ change the output of $f$. This is useful for sensitivity analysis among other things.  In order to isolate these effects to changes in a single input variable, or a single coordinate of an input vector we need some way to encode \emph{one-hot vectors}. For example, if we want to know the derivative of the output array with respect to $x[2]$, we can set $dx$ to be an \emph{one-hot vector} with value $1$ at index $2$ and $0$ elsewhere.  We do this using the \emph{Iverson bracket} notation:
\begin{equation*}
    \kwd{let} dx \keq \gen_{k=-m+1}^{n} \ind{k = 2}\cdot 1.
\end{equation*}
where $\ind{k=2}$ is the indicator function (Iverson bracket), evaluating to $1$ when the predicate inside the brackets is true and $0$ otherwise.

If we plug this one-hot vector into our derivative expression and algebraically simplify, then we arrive at an expression for the sensitivity of convolution to a change in $x[2]$:
\begin{eqnarray*}
    \mathbf{D} f\left(x,\gen_{k=-m+1}^n\ind{k = 2}\cdot 1\right)
    &=& \gen_{i=0}^{n} \ssum_{j=0}^{m} \left(\gen_{k=-m+1}^{n} \ind{k = 2} \cdot 1\right)[i - j]\cdot c[j] \\
    &=& \gen_{i=0}^{n} \ssum_{j=0}^{m} \ind{i-j = 2} \cdot c[j] \\
    &=& \gen_{i=0}^{n} \ssum_j \ind{j = i - 2 \conj 0 \le j < m} \cdot c[j] \\
    &=& \gen_{i=0}^{n} \ind{0 \le i-2 < m} \cdot c[i-2] \\
\end{eqnarray*}

Let $\mathbf{D^T}f(x)$ denote the \emph{adjoint derivative}, which may be defined:
\begin{equation*}
    \mathbf{D^T}f(x,dy) = \left(\mathbf{J}f(x)\right)^T \cdot dy
\end{equation*}
Note that the type of the adjoint reflects this transposition $\mathbf{D^T}f : \R^{n+m} \to (\R^n \to \R^{n+m})$.  Efficient means of computing the adjoint are commonly called \emph{reverse mode} differentiation.

In applications such as optimization or machine learning, we are often interested in finding a minimizing input to some scalar loss (eqv. energy) functions $l : \R^n \to \R$.  Gradient descent (also its variants and other methods entirely) require computing the gradient $\nabla l(x)$ as an inner loop sub-routine.  We can define the gradient $\nabla l : \R^n \to \R^n$ in terms of the adjoint derivative by $\nabla l(x) = \mathbf{D^T}f(x,1)$.

For our example, we can ``deconvolve'' a signal $z : \R^n$ to recover some original $x : \R^n$ by minimizing the following sum-of-squares loss function:
\begin{equation*}
    l(x) = \ssum_{k=0}^{n} \left( f(x)[k] - z[k] \right)^2
\end{equation*}

We could compute $\mathbf{D} l(x): \R^{n+m} \rightarrow \R$ in the same way we did for differentiating $f$ and use those results to compute the gradient, but doing so would be inefficient.  We would need to probe $\mathbf{D} l(x)$ $n+m$ times, each time with a different one-hot vector.  We would waste most of our time needlessly computing on zeros.

Instead, we will directly compute $\mathbf{D^T}f$ by \emph{transposing} the linear function directly (i.e. algebraically).  To do this, we rely on the universal property of adjoints/transpositions.  Namely, for any linear function $\mathbf{D}f(x) : \R^n \to \R^m$, its adjoint $\mathbf{D^T}f(x) : \R^m \to \R^n$ is the unique linear function such that forall $dx : \R^n$, $dy : \R^m$
$$ \inner{dy}{\mathbf{D}f(x,dx)} = \inner{\mathbf{D^T}f(x,dy)}{dx} $$
where $\inner{\cdot}{\cdot}$ denotes the conventional inner product (dot product) for any given vector space.

Equipped with this perspective, we can reduce gradient differentiation to rote algebraic manipulation.  Letting $dy = 1$ as specified by $\nabla l(x) = \mathbf{D^T}l(x,1)$
\begin{align*}
   \inner{1}{\mathbf{D} l(x,dx)}
 = & \inner{1}{\deriv{\ssum_{i=0}^{n}\left(f(x)[i] - z[i]\right)^2}{x\mapsto dx}} \\
 = & \inner{1}{\ssum_{i=0}^{n}2\cdot\left(f(x)[i] - z[i]\right)\cdot\deriv{f(x)}{x\mapsto dx}[i]} \\
 = & \msum_{i=0}^{n}\inner{2\cdot \left(f(x)[i] - z[i]\right)}{\deriv{f(x)}{x\mapsto dx}[i]} \\
 = & \inner{\gen_{i=0}^{n}2\cdot \left(f(x)[i] - z[i]\right)}{\mathbf{D}f(x,dx)} \\
 = & \inner{\mathbf{D^T}f\left(x,~\gen_{i=0}^{n}2\cdot \left(f(x)[i] - z[i]\right)\right)}{dx} \\
\end{align*}
Importantly, note the second to last line here, in which a summation of inner products is converted to an inner product of vectors.  This follows from the definition of inner product.  However, it also illustrates an important principle of adjoint differentiation.  The transposition of a summation (as in the sum-of-squares loss) is the generation of a vector.  Observing the types, this must necessarily be the case, since $l:\R^n \to \R$.

Then, we continue by working out the adjoint derivative of $f$, not just $l$.  We get the following sequence of manipulations (with $dy : \R^n$):
\begin{align*}
 \inner{dy}{\mathbf{D} f(x,dx)}
 = & \inner{dy}{\gen_{i=0}^{n} \ssum_{j=0}^{m} dx[i - j]\cdot c[j]} \\
 = & \msum_{i=0}^{n} \msum_{j=0}^{m} \inner{dy[i]\cdot c[j]}{dx[i - j]} \\
 = & \msum_{i=0}^{n} \msum_{j=0}^{m} \inner{\gen_{k=-m+1}^{n} \ind{k = i - j} dy[i] c[j]}{dx} \\
 = & \inner{\ssum_{i=0}^{n} \ssum_{j=0}^{m} \gen_{k=-m+1}^{n} \ind{k = i - j} dy[i] c[j]}{dx} \\
\end{align*}

The most important rewrite is the second to last equation where we want to rewrite the right-hand side from $dx[i - j]$ to $dx$. To do this we have to introduce an extra inner loop that goes over all elements on the left-hand side, and choose only the elements that matches the index $i - j$. This might seems wasteful at the first glance. However we can eliminate the sum over $i$ by observing that $i = k + j$ from the predicate:
\begin{equation*}
    \ssum_{i=0}^{n} \ssum_{j=0}^{m} \gen_{k=-m+1}^{n} \ind{k = i - j}\cdot dy[i]\cdot c[j] =
    \ssum_{j=0}^{m} \gen_{k=-m+1}^{n} dy[k + j]\cdot c[j] = \gen_{k=-m+1}^{n} \ssum_{j=0}^{m} dy[k + j]\cdot c[j].
\end{equation*}
Thus we arrive at the well-known result that the adjoint of a convolution is a correlation.

If we compose this result with the adjoint differentiation of our loss function, we arrive at the following expression for the gradient $\nabla l(x)$
\begin{align*}
\kwd{let} dy
    &= \gen_{i=0}^n 2 \cdot \left(\left(\ssum_{j=0}^m x[i-j]\cdot c[j] \right)- z[i]\right) \\
\kwd{in}&
    \gen_{i=-m+1}^{n} \ssum_{j=0}^{m} dy[i + j]\cdot c[j]
\end{align*}
which is a pipeline of two stencil operations.  Languages like Halide~\cite{Ragan-Kelley:2013:HLC} are designed to exploit code in this form by making subtle trade-offs between parallelism, recomputation and memory locality.  Depending on the size of the constant $m$ and the hardware targeted, it may make sense to fully fuse these loops or only to partially tile and fuse the loops.  Maintaining the code in a fully functional form makes it far easier for downstream compilation to apply such techniques.

In traditional automatic differentiation literature, reverse-mode differentiation is known to satisfy the \emph{cheap gradient principle}~\cite{Griewank:2008:EDP}, which says that the time complexity of the adjoint derivative $\mathbf{D^T}f(x)$ is the same as the original function $f(x)$. We next show that it is non-trivial for this to hold in a purely functional setting using a different example.

\subsection{Zero Costs}

Consider the function $\textrm{trace}(\textrm{eye}(x))$ of type $\R\to\R$, similar to the examples given in our introduction.  The operators can be defined using our notation as
\begin{align*}
    \textrm{trace}(A) =\ & \ssum_{i=0}^n A[i,i] \\
    \textrm{eye}(x) =\ & \gen_{i=0}^n\gen_{j=0}^n \ind{i\keq j}\cdot x \\
\end{align*}
Using the method outlined in our previous example, we can show that $\mathbf{D^T}\textrm{trace}(A,dx) = \text{eye}(dx)$ and that $\mathbf{D^T}\textrm{eye}(x,dA) = \textrm{trace}(dA)$.

In the folk-proof understanding of the \emph{cheap gradients principle} and in Griewank and Walther~\cite{Griewank:2008:EDP} we assume that (1) cost decomposes \emph{additively} over composition of elementary functions and (2) cost is preserved by transposition/adjoint of those elementary functions.  Taken at face value, the above example violates principle (2) since trace takes $O(n)$ time/memory-accesses, while eye requires $O(n^2)$ writes.  However, the transposition of the two operations together still end up balancing each other out and satisfying the cheap gradients principle.

The argument by Griewank and Walther remains sound because they do not rely on decomposing a program into the purely functional composition of some collection of primitives.  In their imperative setting, the intermediary $A$ matrix must be allocated and zeroed out in the original function and in the adjoint derivative.  Furthermore, the eye function resulting from adjunction of trace (when implemented imperatively) is executed as a loop of $n$ sparse updates ($dA[i,i] \pluseq x$) to an intermediary matrix.

The argument is not sound in a functional setting without recourse to these sparse mutating updates.  Furthermore, there is no single, obvious way to impose a cost model on \emph{writes} and \emph{reads} in functional code.  For instance, the previously discussed sum-of-squares loss function may be written as $\ssum_{i=0}^n x[i]*x[i]$, or it may be written equally well as
$$ \kwd{let} y = \gen_{i=0}^n x[i]\cdot x[i] \kwd{in} \ssum_{i=0}^n y[i]$$
which is more analogous to a notation like $\textrm{sum}(\textrm{map}((\lambda a. a\cdot a), x))$.  Both forms undoubtedly perform $n-1$ additions and $n$ multiplications, but one may credibly argue that the former only reads from memory $n$ times and writes once, while the latter reads $2n$ times and writes $n+1$ times.  Other accounting schemes may even place the count at $3n$ reads since $x[i]$ is being read ``twice.''

In order to avoid these issues, we propose to break from Griewank and Walther by only counting arithmetic operations towards the cost.

If we revisit the scalar case ( i.e. without $\gen$ or $\ssum$ constructs ) and strictly count arithmetic operations, we can find some programs whose adjoint has substantially more additions than the original program has.  For instance, consider this pathological scalar function $f(x)$, full of un-used intermediary copies of the input (aka. ``dead code'')
\begin{align*}
    \kwd{let} z_0 &= x \kwd{in} \\
    \kwd{let} z_1 &= x \kwd{in} \\
    &\vdots \\
    \kwd{in}\ & 2\cdot x
\end{align*}
When fed na\"ively to reverse-mode automatic differentiation, the result is the following program
$$ 2\cdot dy + 0 + 0 + \cdots $$
which has $n$ additions even though the original function had only $1$ multiplication and $0$ additions.
(If we take a sufficiently na\"ive reverse-mode derivative of this new function, we will arrive back at something like our original function, since both are linear.)

From this we can observe a general principle: the adjoint of dead code is zeros, and the adjoint of a zero is dead code.  We can also observe the well known general principle that the adjoint of \emph{fan-out} (i.e. the repeated re-use of the variable $x$) is summation, and dually the adjoint of summation is fan-out.  From a functional perspective, it seems odd to impose a cost on fan-out, which helps to create the preceding problem.

For instance, consider this linear program and its corresponding compute DAG:
\begin{figure*}[h!]
\begin{subfigure}{0.3\textwidth}
    \begin{align*}
        \kwd{let} a &= 2\cdot x + 3\cdot y \kwd{in} \\
        \kwd{let} b &= 4\cdot x + a \kwd{in} \\
        \kwd{let} z &= 8\cdot x \kwd{in} \\
        \kwd{let} w &= 5\cdot a + 6\cdot b \kwd{in} \\
        &(z,w)
    \end{align*}
\end{subfigure}
\begin{subfigure}{3in}
    \includegraphics[width=3in]{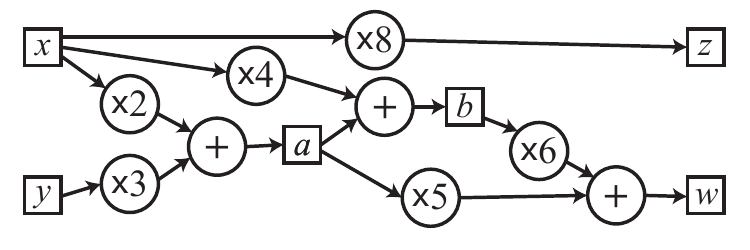}
\end{subfigure}
\end{figure*}

It is common to think of the transpose or adjunction of this code as simply ``the same graph with arrows turned backwards.''  (In a suitable DAG formalism this idea holds.)  However, notice that when we ``turn the arrows backwards'' fan-out and fan-in are reversed, leading to non-local motion of addition operations:
\begin{figure*}[h!]
\begin{subfigure}{0.3\textwidth}
    \begin{align*}
        \kwd{let} b &= 6\cdot w \kwd{in} \\
        \kwd{let} a &= 5\cdot w + b \kwd{in} \\
        \kwd{let} x &= 2\cdot a + 4\cdot b + 8\cdot z \kwd{in} \\
        \kwd{let} y &= 3\cdot a \kwd{in} \\
        &(x,y)
    \end{align*}
\end{subfigure}
\begin{subfigure}{3in}
    \includegraphics[width=3in]{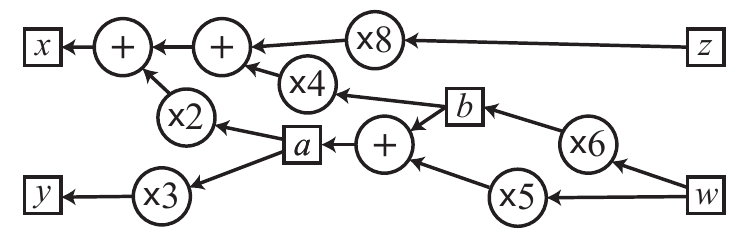}
\end{subfigure}
\end{figure*}

Somewhat miraculously, the overall number of additions has been preserved despite this non-local motion.  We can build an argument for the \emph{global} preservation of additions by counting the number of edges, less the number of (non-input) nodes: $A = \#E - \#N$.  (note that scaling has been represented as a unary operation here.)  This is precisely the count of the number of additions required.  By simply adjusting for different numbers of input/output nodes, we can observe that this quantity is preserved whenever we flip the edges.

This argument fails whenever we have dead-code or $0$s, because (from the DAG point of view) these look like spurious additional outputs and inputs.  However dead-code elimination and constant propagation can always remove all of these complications.

In the tensor setting, zeros and dead-code become ubiquitous, as sparse access and sparsity of tensors.  As the example of the trace function shows, sparsity in this sense arises even in dense tensor data.  In order to extend our idea for a scalar argument to tensor data, we will need to handle sparsity explicitly, which we do via the indicator function/Iverson bracket $\ind{\cdot}$.

As a representative example, consider the simple program with input $x : \R^n$
$$ x[0] + x[2] + x[3] + x[4] + \cdots $$
where we have oddly skipped summing over entry $x[1]$.  This function takes $O(n)$ additions to compute.  Na\"ively taking the adjoint, we have 
$$\left(\gen_{i=0}^n \ind{i\keq 0}\cdot dy\right) + \left(\gen_{i=0}^n \ind{i\keq 2}\cdot dy\right) +
  \left(\gen_{i=0}^n \ind{i\keq 3}\cdot dy\right) + \left(\gen_{i=0}^n \ind{i\keq 4}\cdot dy\right) + \cdots$$
which would appear to take $O(n^2)$ additions, primarily of zeros.  However, if we keep track of sparsity via these indicator functions and propagate that information, we can instead focus on computing
$$\gen_{i=0}^n\left( \ind{i\keq 0}\cdot dy + \ind{i\keq 2}\cdot dy +
                     \ind{i\keq 3}\cdot dy + \ind{i\keq 4}\cdot dy + \cdots \right)$$
This is essentially a synonym for the tensor literal
$$ [dy, 0, dy, dy, dy, \hdots] $$
If we implement tensors in a sparsity-aware manner (e.g. merging sparse tensor representations as TACO~\cite{kjolstad2017tensor} does) or compile that sparsity structure into the code, then we can eliminate these spurious additions.

In our cost model, we consequently say that additions of the form $\ind{p_0}\cdot e_0 + \ind{p_1}\cdot e_1$ only counts as an addition when both $p_0$ and $p_1$ are true.  Similarly, a summation over a condition which is only true exactly once (e.g. $\ssum_{i=0}^n\ind{i=\cdots}\cdot e$) doesn't perform any additions because it may be compiled into pure indexing arithmetic. (which we do not count, following Griewank, Walther and all prior automatic differentiation cost models)

\subsection{Normalizing Code}
\label{sec:ex-norm-code}

Carefully handling intermediate values (i.e. $\kwd{let}$ bindings) is also crucial for preserving the time complexity.
Consider the previous deconvolution example, with the input signal $x : \R^{(n+m-1)}$ held constant and the weight kernel $w : \R^m$ varying:
\begin{align*}
    f(x;w) &= \gen_{i=0}^{n} \ssum_{j=0}^{m} x[i - j]\cdot w[j] \\
    l(x,z;w) &= \ssum_{i=0}^{n} (f(x;w)[i] - z[i])^2
\end{align*}
Taking the adjoint as before yields the following loss-gradient:
\begin{align*}
    \mathbf{D^T} f(x;w,dy) &= \gen_{j=0}^m \ssum_{i=0}^{n} x[i-j] \cdot dy[i] \\
    \nabla l(x,z;w) &= \kwd{let} dy = \gen_{i=0}^{n} 2\cdot\left(
                                        \left(\ssum_{j=0}^{m} x[i-j]\cdot w[j]\right) - z[i]\right) \\
                  &\phantom{=}\ \ \kwd{in}\ \mathbf{D^T} f(x;w,dy) \\
\end{align*}

Given a collection of signals before and after an unknown convolution, we can attempt to fit a linear model to that convolution using this loss gradient.  Let $x : \R^{B\times(n+m-1)}$ be the set of $B$ inputs, and $z : \R^{B\times n}$ be the set of $B$ outputs.  Without loss of generality, this may be either the entire data set or a single \emph{batch}.  Then the loss-gradient for the batch is
$$ \everymath={\displaystyle}
\nabla l(x,z;w) = \ssum_{k=0}^{B} \left(\begin{array}{rl}
\kwd{let} dy &=
    \gen_{i=0}^{n} 2\cdot\left(\left(\ssum_{j=0}^{m} x[k,i-j]\cdot w[j]\right) - z[k,i]\right) \\
&\kwd{in}\ \gen_{j=0}^m \ssum_{i=0}^{n} x[k,i-j] \cdot dy[i]
\end{array}\right) $$

Observe that $\kwd{let}$ bindings get nested within ``loops'' here, which is a very common phenomenon.  While such nesting can be desirable for efficient execution, it also impedes program analyses and transforms.  For instance, adjoint differentiation cannot generally preserve this structure without again relying on adding an impure $\pluseq$ operation to the language.  For this reason, we find it necessary to normalize tensor code by \emph{flattening} these structures.  For the above program, the \emph{let-lifting} transformation/normalization yields
$$ \everymath={\displaystyle}
\nabla l(x,z;w) = \begin{array}{rl}
\kwd{let} dy &=
    \gen_{k=0}^{B}\gen_{i=0}^{n} 2\cdot\left(\left(\ssum_{j=0}^{m} x[k,i-j]\cdot w[j]\right) - z[k,i]\right) \\
&\kwd{in}\ \ssum_{k=0}^{B}\gen_{j=0}^m \ssum_{i=0}^{n} x[k,i-j] \cdot dy[k,i]
\end{array} $$

For the Tensor Language we describe here, this transform can always be performed without increasing the operation-count cost of a program.  As a result, we can always normalize code in this manner in order to take adjoint derivatives.  Furthermore, this kind of transformation can expose more scheduling, optimization and parallelization opportunities to downstream systems.  For instance, the above convolutions are now batched and can be targeted to high-performance matrix-matrix multiply hardware using the image-to-column transformation strategy~\cite{im2col}.

\paragraph{Overview.} In the following, after discussing the relation to prior works, we first define our Tensor Language, then we details our cost model. Motivated from the cost model, we develop transformation to convert tensor programs into a normal form, and explain the differentiation and transposition rules.

\section{Relation to Prior Work}

\subsection{Automatic differentiation}

\paragraph{Deep learning systems.}
Existing deep learning tensor frameworks~\cite{Abadi:2015:TLM, Paszke:2019:PIS, Bergstra:2010:TCG} can be seen as domain specific languages that are built around the chain rule as a modularity principle for composing bulk data-parallel operators/layers, such as convolution and matrix multiplication.
This allows them to achieve high performance for dense, multi-layer neural networks thanks to the highly-optimized manually-written operators and their high arithmetic intensity.
Unfortunately, if a desired bulk operation is not already provided, users must develop their own~\cite{Barham:2019:MLS, Li:2018:DPI}, or rely on inefficient scalar automatic differentiation systems that does not reason about tensor code.

In particular, deep learning systems do not support higher-level operations on code, (e.g. as discussed in \S\ref{sec:ex-norm-code}) which requires all primitives to have built-in batching dimensions, etc.  (JAX~\cite{Bradbury:2018:JCT} is a notable exception in this regard.)  As a result, the kind of chain-rule/composition considered is limited strictly to a DAG of operators.  This limitation is likewise present in recent functional language work on AD~\cite{Elliott:2018:SEA, Brunel:2019:BST, Abadi:2019:SDP, Sherman:2020:CSD}, as we will soon discuss.

\paragraph{Reverse-mode automatic differentiation.}
\emph{Derivative transposition}, the \emph{adjoint derivative}, or \emph{reverse-mode} automatic differentiation, was first derived by Linnainmaa~\cite{Linnainmaa:1970:RCR} and was invented independently several times later~\cite{Griewank:2012:WIR}. Our adjoint derivative rules are derived from the \emph{universal property} of adjoints/transpositions of linear functions rather than as a procedural organization of how partial derivatives are computed.  This stylistic novelty allows us to quickly certify that our adjoint-derivative rules are correct by construction.

In this respect, we push forward Conal Elliott's categorical point of view~\shortcite{Elliott:2018:SEA} while exploiting the bilinearity of the inner-product more directly.  He observed that unlike the forward/total-derivative, implementing reverse mode requires specialization from the more abstract setting of Cartesian closed categories to biproduct categories.  In particular this requires the choice of an inner product, because the mapping between a vector space and its dual is a non-canonical isomorphism.

\paragraph{Wengert tapes.}
To compute adjoint derivatives of imperative programs with loops and complex control flows, most systems rely on a tape/list data-structure to record values from the ``forward pass'' which must then be unwound upon reaching the end of execution of the function to be differentiated.  While this structure is commonly known as a Wengert tape, referring to the original publication on automatic-differentiation~\cite{Wengert:1964:SAD}, it appears this is a mis-attribution, as Wengert's method was limited to forward differentiation---it was left to later authors to innovate this mechanism.  The tape strategy can also be used for statically compiling derivative programs that instrument loops to construct the tape.  Without extra care and optimization, this strategy produces a lot of unnecessary tape-management code, as well as large amounts of pointer indirection into the heap, both of which impede downstream optimizations.  Still, this remains the most popular and generalizable approach to reverse-mode automatic differentiation.  Systems such as ADIFOR~\cite{Bischof:1992:AAD}, ADrien~\cite{Villard:1999:AIA}, OpenAD~\cite{Utke:2008:OMO}, Tapenade~\cite{Hascoet:2013:TAD}, Zygote~\cite{Innes:2018:DUA}, RelayIR~\cite{roesch2018relay}, DiffTaichi~\cite{hu2019difftaichi} all use variants of this strategy, and employ a side-effecting $\pluseq$ accumulation to the references stored in or referenced by the tape.

Interestingly, even the seeming functional works on automatic differentiation employ a similar strategy using Wengert's tape that involves side effects. Stalingrad~\cite{Pearlmutter:2008:RAF} adds an imperative mutation command to an otherwise purely-functional language. The necessity of this strategy is precipitated by the need to handle general \emph{closures}.  In the reverse mode version of a function, the ``feedback'' to variables closed over must be rendered as accumulation to a dynamically determined variable location captured in the closure. Similarly, Lantern~\cite{Wang:2019:DDP} employs the $\pluseq$ mutation and refers them as the \emph{destination-passing style}. As previously mentioned, other functional works focus on simpler constructs and do not handle loops~\cite{Brunel:2019:BST, Abadi:2019:SDP, Sherman:2020:CSD}.

Li et al.~\cite{Li:2018:DPI}, when developing an automatic differentiation extension to the Halide language, noted that the $\pluseq$ scattering mutation can hurt parallelism. They proposed a scatter-to-gather conversion to improve the performance. Huckelheim et al. later proposed a similar technique specialized to stencil code~\cite{Huckelheim:2019:ADA, Huckelheim:2019:TPA}.  Unfortunately, neither of these conversions guarantee that all $\pluseq$ operations are removed, forcing incorporating of imperative language constructs. 

In contrast, we focus on a more constrained form of loops ($\gen$ and $\ssum$). This allows us to avoid the use of tapes.  To avoid the scattering operations, we keep track of the sparsity structures of the tensor expressions and propagate the derivatives by carefully inverting the sparsity links. As a result, all of our constructs and their derivatives are intrinsically parallel.

\paragraph{Optimizing and simplifying derivative code.}

Transforming the derivative code to achieve higher performance can be seen as the task of an optimizing compiler. Earlier works on automatic differentiation use different greedy strategies~\cite{Griewank:1991:OCJ, Naumann:1999:ECJ, Guenter:2007:ESD, Wang:2016:EPE} to factor the common subexpressions using Bauer's formula~\cite{Bauer:1974:CGR}. Naumann further proves that the factorization leading to the minimal calculation is NP-complete~\cite{Naumann:2008:OJA}. More recently, Shaikhha et al.~\cite{Shaikhha:2019:EDP} propose a set of rewrite rules for optimizing derivative code. Laue et al.~\cite{Laue:2018:CHO} also collect a set of tensor calculus rules for simplifying tensor derivative.

Many works treat the functions resulting from differentiation with special types such as tangent bundles~\cite{Brunel:2020:BST} or linear function types~\cite{Elliott:2009:BD}.  By contrast, we do not make any special distinctions between this linear derivative code and general (non-linear) numeric code.  As a result, we are free to optimize the linear, polynomial and non-linear parts of our programs jointly, without distinction as to whether they were written by hand, obtained from derivative transformations, or came from some repeated interleaving thereof.

\subsection{Tensor Languages and Optimizing Compilers}

Domain specific languages with tensors or multi-dimensional arrays as first-class constructs are common in a wide range of fields including machine learning, physics simulation, and image processing.  Crucially, such languages allow for effective exploitation of data parallelism.

Deep learning frameworks are one example of these domain specific languages, which we briefly introduced in the previous subsection. JAX~\cite{Bradbury:2018:JCT} adapts this idea to compile and differentiate NumPy programs more generally, using the XLA~\cite{xla} backend to optimize the resulting programs and target acclerators like Google's Tensor Processing Unit.  Other works focus on the algebraic simplification of these tensor frameworks (e.g.,~\cite{Jia:2019:TOD}).  It is important for such systems to include rewrite rules that exploit the linearity of the differentiation (such as the derivative transposition we use in this paper), otherwise it is difficult to provide performance guarantee on larger-scale computation.

Halide~\cite{Ragan-Kelley:2012:DAS, Ragan-Kelley:2013:HLC} posits separate schedules as a rewrite system to optimize dense tensor code. Polyhedral compilers (e.g.,~\cite{Bondhugula:2008:PAP, Grosser:2012:PPP, vasilache2018tensor}) also transform affine loop nests for optimization. TACO~\cite{kjolstad2017tensor}, on the other hand, treats sparse-tensor computations. However, it focuses almost exclusively on linear algebra problems, and remains very focused on individual kernels. Taichi~\cite{hu2019taichi} focuses on optimizing spatially coherent sparse computation. These rewrite systems and optimizing compilers can potentially be used together with our compiler during differentiation. In this paper we focus on transformations for generating efficient, parallel derivative code and the formal arguments that provide performance guarantees.

At a slightly higher level, another group of domain specific languages/systems focuses on solving specific loss optimization problems~\cite{heide2016proximal, ceres-solver, Devito:2017:ODS, diamond2016cvxpy, agrawal2018rewriting}. These systems show that reasoning about optimizing code for the full loss-optimization problem is crucial for high performance code generation.  For instance, Opt~\cite{Devito:2017:ODS} computes a fusion between the derivative and adjoint derivative to yield a simplified, matrix-free Jacobian-transpose-Jacobian-vector product that is compiled for GPU, and injected into the inner loop of a conjugate gradient solver inside of a Gauss-Newton optimizer.  Our work can be thought of as a first step towards generalizing such systems.

\section{Language Definition}
\label{sec:lang-def}

\subsection{Definition and Evaluation}

\begin{figure}[t]
\begin{tabular}{rcll}
\toprule
$x$  &       &                                      & variable \\
$n$  &       &                                      & size variable \\
$R$  &       &                                      & relation variable \\
\midrule
$e$  & $::=$ & $x$                                  & variable name \\
     &  $|$  & $c$                                  & constant number \\
     &  $|$  & $e_0 + e_1$                          & scalar addition \\
     &  $|$  & $e_0 \cdot e_1$                      & scalar multiplication \\
     &  $|$  & $f(e_0,\hdots)$                      & black-box scalar function \\
     &  $|$  & $(e_0 , e_1)$                        & pair construction \\
     &  $|$  & $\pi_0~e$ $~|~$ $\pi_1~e$            & pair projection \\
     &  $|$  & $\gen_{i=0}^n e$                     & tensor generation \\
     &  $|$  & $\ssum_{i=0}^n e$                    & tensor summation \\
     &  $|$  & $e[a]$                               & tensor access \\
     &  $|$  & $\ind{p}\cdot e$                     & Iverson bracket (indicator) \\
     &  $|$  & $\kwd{let} x = e_0 \kwd{in} e_1$     & let-bindings \\
\midrule
$p$  & $::=$ & $a_0 < a_1$ $~|~$ $a_0 \leq a_1$     & comparison predicates\\
     &  $|$  & $a_0 \keq a_1$                       & \\
     &  $|$  & $R(a_0,\hdots)$                      & relational (data) predicates\\
     &  $|$  & $p_0 \conj p_1$ $~|~$ $p_0 \disj p_1$ & conjunction and disjunction\\
     &  $|$  & $\exquant_{i=0}^n p$                  & existential quantification\\
\midrule
$a$  & $::=$ & $i$                                  & index variable name \\
     &  $|$  & $n$                                  & size variable name \\
     &  $|$  & $k$                                  & integer constant \\
     &  $|$  & $a_0 + a_1$                          & integer addition \\
     &  $|$  & $k \cdot a$                          & integer scaling \\
\midrule
$T$  & $::=$ & $\mathbb{R}$                         & number type \\
     &  $|$  & $T_0\times T_1$                          & pair type \\
     &  $|$  & $[n]T$                               & tensor type \\
\bottomrule
\end{tabular}
\caption{Grammar for Core Tensor Language}
\label{fig:atl_grammar}
\end{figure}

In order to study the simplification and compilation of functional tensor languages, we define A Tensor Language (Figure \ref{fig:atl_grammar}).  This language is not Turing-complete or fully featured.  It can be used in DSLs for specifying energy functions to optimize, or it can be used as an inner-loop fragment in some larger language.  It represents a kind of ``basic block'' sub-language amenable to the simplification techniques we will explore.
See the examples for a more intuitive introduction (\S\ref{sec:example-intro}).

\begin{figure}[t]
\begingroup
\newcommand{\E}[2]{\EvalExpr{#1}{#2}}
\arraycolsep=1.8pt
\begin{framed}
\begin{tabular}{rcll}
$v$  & $::=$ & $c$                                  & scalar values \\
     &  $|$  & $(v_0,v_1)$                          & pair values \\
     &  $|$  & $[v_0,\hdots,v_{n-1}]$               & array values \\
$0_\R$          & $=$ & $0$                         & zero shapes   \\
$0_{(T_0,T_1)}$ & $=$ & $(0_{T_0},0_{T_1})$     \\
$0_{[n]T}$      & $=$ & $[0_T,\hdots,0_T]$      \\
\end{tabular}
\vspace{10pt}
\hrule
\begin{align}
\E{x}{\sigma}             &= \sigma(x)                              \\
\E{c}{\sigma}             &= c                                      \\
\E{e_0 + e_1}{\sigma}     &= \E{e_0}{\sigma} + \E{e_1}{\sigma}      \\
\E{e_0 \cdot e_1}{\sigma} &= \E{e_0}{\sigma} \cdot \E{e_1}{\sigma}  \\
\E{f(e_0,\hdots)}{\sigma} &= f(\E{e_0}{\sigma},\hdots)              \\
\E{(e_0,e_1)}{\sigma}     &= \left(\E{e_0}{\sigma},\E{e_1}{\sigma}\right) \\
\E{\gen_{i=0}^n e}{\sigma}  &= \left[~ \E{e}{\sigma\msub{i\mapsto0}},
                              \hdots,
                              \E{e}{\sigma\msub{i\mapsto n-1}} ~\right]  \\
\E{\ssum_{i=0}^n e}{\sigma} &= \E{e}{\sigma\msub{i\mapsto0}} + \cdots +
                              \E{e}{\sigma\msub{i\mapsto n-1}}           \\
\E{\kwd{let} x \keq e_0
   \kwd{in} e_1}{\sigma}  &= \E{e_1}{\sigma\msub{x\mapsto
                                                 \E{e_0}{\sigma}}}
\end{align}
\begin{equation}
\begin{mathprooftree}
  \AxiomC{$\E{e}{\sigma} = (v_0,v_1)$}
  \UnaryInfC{$\E{\pi_0~e}{\sigma} = v_0$}
\end{mathprooftree}
\end{equation}
\begin{equation}
\begin{mathprooftree}
  \AxiomC{$\E{e}{\sigma} = [v_0,\hdots,v_{n-1}]$}
  \AxiomC{$\E{a}{\sigma} = k$}
  \BinaryInfC{$\E{e[a]}{\sigma} = v_k$}
\end{mathprooftree}
\end{equation}
\begin{equation}
\begin{mathprooftree}
  \AxiomC{$\E{p}{\sigma} = \kwd{true}$}
  \UnaryInfC{$\E{\ind{p}\cdot e}{\sigma} = \E{e}{\sigma} $}
\end{mathprooftree}
%
\begin{mathprooftree}
  \AxiomC{$\E{p}{\sigma} = \kwd{false}$}
  \UnaryInfC{$\E{\ind{p}\cdot e}{\sigma} = 0_{\textrm{type}(e)} $}
\end{mathprooftree}
\end{equation}
\end{framed}
\caption{Operational Semantics for Core Tensor Language}
\label{fig:semantics}
\endgroup
\end{figure}

\begin{figure}[t]
\begingroup
\newcommand{\E}[2]{\EvalExpr{#1}{#2}}
\newcommand{\op}{\langle op \rangle}
\arraycolsep=1.8pt
\begin{framed}
\begin{align}
\E{i}{\sigma}           &= \sigma(i)                            \\
\E{n}{\sigma}           &= n                                    \\
\E{k}{\sigma}           &= k                                    \\
\E{a_0 + a_1}{\sigma}   &= \E{a_0}{\sigma} + \E{a_1}{\sigma}    \\
\E{k \cdot a}{\sigma}   &= k \cdot \E{a}{\sigma}                \\
\E{a_0 \op a_1}{\sigma} &= \E{a_0}{\sigma} \op \E{a_1}{\sigma}
                           \hspace{1cm} where \op \in \{<,\leq,\keq\} \\
\E{p_0 \op p_1}{\sigma} &= \E{p_0}{\sigma} \op \E{p_1}{\sigma}
                           \hspace{1cm} where \op \in \{\conj,\disj\} \\
\E{R(a_0,\hdots)}{\sigma}&= R(\E{a_0}{\sigma},\hdots) \\
\E{\exquant_{i=0}^n p~}{~\sigma} &= \E{p}{\sigma\msub{i\mapsto 0}} \disj \cdots \disj
                                    \E{p}{\sigma\msub{i\mapsto n-1}}
\end{align}
\end{framed}
\caption{Operational Semantics for Predicates and Affine Indices}
\label{fig:idxpred-semantics}
\endgroup
\end{figure}

Given an expression $e$ with free variables $FV(e)$, and an environment $\sigma$ mapping those free variables to values, we define the evaluation (Figures \ref{fig:semantics} \& \ref{fig:idxpred-semantics}) of $e$ denotationally or (big-step) operationally as $\EvalExpr{x}{\sigma}$.  Note that we use \textcolor{meta}{red} to help distinguish meta-linguistic transformation constructs, such as evaluation.  Operations on the right-hand-side of evaluation rules (e.g. $+$, $\cdot$) are assumed to be evaluated in their respective value-domains (real or rational numbers, with floating-point approximations used in practice).

We rely on a standard environment formalism, in which $\sigma\msub{ x\mapsto v }$ represents environment extension with a new mapping (shadowing all pre-existing mappings to its left).  Singleton environments are simply written $\msub{ x \mapsto v}$, while empty environments are written $\msub{}$.  When we get to program transformations, it will be useful to write $\msub{x \mapsto v}e$ to mean the expression $e$ with the specified substitution applied.  This last form will allow us to use the identity $\EvalExpr{\msub{x\mapsto v}e}{\sigma} = \EvalExpr{e}{\sigma\msub{x\mapsto v}}$

Typing rules follow as expected, with ``index types'' distinguished grammatically in our presentation.  The $\gen_{i=0}^n e$ operation has\footnote{Our unfamiliar convention of prefixing the size of an array type nicely resolves a syntactic paradox for multi-dimensional arrays:  Let $x$ be an $n$ by $m$ matrix.  If x has type $(\R[n])[m]$, then $x[i] : \R[n]$, so in the expression $x[i][j]$ the index $i$ must counter-intuitively correspond to the $m$ dimension.  If instead $x : [n]([m]\R)$, the problem is neatly resolved.} type $[n]T$ when $e : T$.  For simplicity of analysis, we consider $+$, $\cdot$, big summation, and all black-box functions to be well-formed only when applied to scalar operands\footnote{one important consequence is that we do not treat some special-purpose bulk-tensor operations that are not representable in our language, such as FFTs}.  (Other forms of addition and multiplication may be de-sugared to this)  Importantly, we consider multiplication by an Iverson-Bracket to be well-formed over any expression type, which helps when expressing simplification rules.

Because only scalars, tensor and pair types are allowed, we can elide a long discussion of typing rules.  However, for an expression with free variables to be well-typed, we must assert types for all of those free variables.  Similarly, we may choose to add type-annotations to let-bindings, as in $\kwd{let} x : T \keq e_0 \kwd{in} e_1$.  We assume that such annotations are inferred for all programs.

\subsection{Expressions Denote Functions}
\label{sec:expressions-are-functions}

Let $e$ be a well-formed and well-typed expression with free-variable types $FVT(e) = \{ x_0 : T_0, \hdots, x_k : T_k\}$.  Then $e$ has some ``return'' type $e : T$ and we may interpret $e$ as a total function $(T_0\times \cdots\times T_k)\to T$.  Furthermore, by associating a finite-dimensional vector space over $\R$ to every type in our language, we may interpret $e$ as isomorphic to some function $f : \R^{n_I} \to \R^{n_O}$ for some positive integers $n_I$ and $n_O$.  To do this, let $V(T)$ denote the vector space corresponding to a type $T$.  $V(\R) = \R$, the $1$-dimensional vector space; $V([n]T) = \R^n\otimes V(T)$ where $\otimes$ denotes a tensor-product vector space; and $V(T_0 \times T_1) = V(T_0) \oplus V(T_1)$ denotes a direct-sum of vector spaces.

Note that if we make no use of the black-box scalar function escape hatch to extend our language, then the class of functions we treat is simply multi-variate polynomials.  If all such primitive scalar functions are smooth, then we are working in the space of smooth functions.  In this way we defer any questions about the analytic differentiability of functions to the allowed primitives, including issues arising from discontinuities.  Without loss of generality we will assume that multi-dimensional derivatives are well defined and all definitions thereof coincide.  (For instance, we will not worry about different directional derivatives approaching cusps from different directions.)

\subsection{Cost-Model}
\label{sec:cost-model}

\begin{figure}[t]
\begingroup
\newcommand{\E}[2]{\CostExpr{#1}{#2}}
\arraycolsep=1.8pt
\begin{framed}
\begin{align}
\E{x}{\sigma}             &= 0                                      \\
\E{c}{\sigma}             &= 0                                      \\
\label{rule:cost-def-binadd}
\E{(\ind{p_0}\cdot e_0) + (\ind{p_1}\cdot e_1)}{\sigma}
                          &= \E{\ind{p_0}\cdot e_0}{\sigma} + \E{\ind{p_1}\cdot e_1}{\sigma}
                             + \left[\E{p_0 \conj p_1}{\sigma}\right] &(*)\\
\E{e_0 \cdot e_1}{\sigma} &= 1 + \E{e_0}{\sigma} + \E{e_1}{\sigma}  \\
\E{f(e_0,\hdots)}{\sigma} &= 1 + \E{e_0}{\sigma} + \cdots           \\
\E{(e_0,e_1)}{\sigma}     &= \E{e_0}{\sigma} + \E{e_1}{\sigma}      \\
\E{\pi_0 e}{\sigma}       &= \E{e}{\sigma}                          \\
\E{\gen_{i=0}^n e}{\sigma}  &= \sum_{\hat{\imath}=0}^n
                                    \E{e}{\sigma\msub{i\mapsto \hat{\imath}}} \\
\label{rule:cost-def-bigsum}
\E{\ssum_{i=0}^n \ind{p}\cdot e}{\sigma} &= \left(\sum_{\substack{
                                                     \hat{\imath}=0 \\
                                                     \EvalExpr{p}{\msub{i\mapsto \hat{\imath}}\sigma}=\kwd{true}
                                                   }}^n
                                    \left(1 + \E{e}{\sigma\msub{i\mapsto \hat{\imath}}}\right)
                                    \right) - \left[ \EvalExpr{\exquant_{i=0}^n p}{\sigma} \right]
                                    &(**)\\
\E{e[a]}{\sigma}          &= \E{e}{\sigma}                          \\
\E{\kwd{let} x \keq e_0
   \kwd{in} e_1}{\sigma}  &= \E{e_0}{\sigma} + \E{e_1}{\sigma} \\
\E{\ind{p}\cdot e}{\sigma} &= 0, \hspace{22pt} \textrm{ where } \EvalExpr{p}{\sigma}=\kwd{false}\\\
\E{\ind{p}\cdot e}{\sigma} &= \E{e}{\sigma}, \textrm{ where } \EvalExpr{p}{\sigma}=\kwd{true}
\end{align}
\raggedright
\textit{On rules marked $*$ and $**$, the square brackets on the right represent an indicator function, adjusting the count by $1$ (upwards or downwards) when the noted condition occurs.} \\
$*$  \textit{Where there is no predicate nested immediately under an addition, the rule is applied s.t. } $p = \kwd{true}$. \\
$**$  \textit{A composite of multiple big-summations followed by a predicate is handled this way.  A summation without a predicate is equivalent to this rule with } $p = \kwd{true}$.
\end{framed}
\caption{(Work) Cost Model for A Tensor Language}
\label{fig:cost-model}
\endgroup
\end{figure}

For the purposes of establishing Griewank-style bounds on the (temporal) overhead introduced by differentiation, we define a cost model (Figure \ref{fig:cost-model}) that counts the number of scalar additions, multiplications, and black-box functions executed---but does not attempt to model memory accesses.  Like with evaluation, we use an environment $\sigma$ to store index variable bindings.  The (work) cost of an expression $e$ is thus given by $\CostExpr{e}{\sigma}$.  This cost does not come close to accurately accounting for data movement, memory-locality, vector-instructions, and any number of other considerations, but the cost is still a reasonable proxy for the benefits of algebraic simplification; especially for asymptotic purposes.

Perhaps more concerning, we do not count any additions, multiplications, or comparisons performed on \emph{index-values}.  This choice reflects existing cost-analyses, such as the one presented by Griewank and Walther~\cite{Griewank:2008:EDP}, which is performed relative to straight-line scalar-valued numeric code, stripped of control flow constructs.  However, this can be thought of as a means of analyzing basic blocks of code, and extended to a broader class of programs with other existing or undiscovered techniques.

Crucially, our cost model incorporates short-circuiting behavior based on sparsity captured by Iverson-Bracket predication.  For instance, simple binary addition (rule \ref{rule:cost-def-binadd} $*$) is defined to have zero cost whenever one of the two arguments is $0$ due to the predicate evaluation.  These sorts of additions are amenable to being compiled away or accelerated via sparse data structures.  Similarly, big summation (rule \ref{rule:cost-def-bigsum} $**$) is defined so that only additions of non-zero iterations count.  When the predicate $p$ is an affine predicate, this is achievable by using the predicate to narrow the loop-iteration bounds and/or strides.  When the predicate $p$ is data-driven, representations like CSR matrices can likewise meet this cost-specification.  And most significantly, when $p$ is true for exactly one iteration, the summation loop can be eliminated entirely from compiled code, resulting in no cost overhead.  This is because in all these cases, we must adjust cost downwards to reflect that a summation of $n$ numbers only requires $n-1$ additions.  This may seem inconsequential, but correctly captures the frequent case where a summation is over exactly one value, and the loop can be eliminated altogether.  Note that multiple big summations (e.g. $\ssum\ssum\ind{p}\cdots$) are assigned cost jointly, using the summation-over-predicate rule.

\section{Normal Forms}
\label{sec:normal-forms}

In this section, we will explain how to convert arbitrary, well-formed tensor programs (\S\ref{sec:lang-def}) into a particular kind of \emph{tensor single static assignment} normal form, via a process of four rewriting passes.  Each pass is guaranteed to preserve observational equality of the program, and to not increase the cost of running the program.

Formally, we define a normalization pass as a function $\gamma: terms \to terms$.  We present normalization passes as collections of directed re-write rules\footnote{In practice, these are translated into structurally-recursive passes over ASTs or DAGs.}.  This presentation choice makes it easier to analyze the correctness and properties of passes.

For each pass that transforms a program $e$ into a program $\gamma(e)$, we want to sketch out proofs of the following properties:
\begin{description}
    \item[correctness] For all well-formed $e$ conforming to the precondition normal form for the pass, $e$ and $\gamma(e)$ are observationally equivalent; i.e. $\forall e,\sigma: \EvalExpr{e}{\sigma} = \EvalExpr{\gamma(e)}{\sigma}$.
    \item[normality] For all well-formed $e$ conforming to the precondition normal form for the pass, $\gamma(e)$ conforms to some postcondition normal form.
    \item[cost-neutrality] There is some constant $c$, s.t. for all well-formed $e$ conforming to the precondition normal form for the pass, $\gamma(e)$ has ``the same cost'' as $e$; i.e. $\exists c \forall e,\sigma: c\cdot\CostExpr{e}{\sigma} \geq \CostExpr{\gamma(e)}{\sigma}$.  (Often we can more stringently specify that $c=1$)
\end{description}

A complete treatment would also want to argue for \emph{confluence} (and termination) of our rewrite systems.  We skip those arguments since they ultimately follow from our choice to use syntax-directed rewrite systems, rather than from interesting properties of the subject we are studying.

\subsection{Let-Lifting}

\begin{figure}[t]
\begingroup
\newcommand{\lin}[3]{\kwd{let} #1 \keq #2 \kwd{in} #3}
\arraycolsep=1.8pt
\begin{framed}
\textit{(assuming no re-use of variable names)}
\begin{align}
\left(\lin{x}{e_0}{e_1}\right) + e_2
                          &\rewritesto  \lin{x}{e_0}{e_1 + e_2}               \\
\left(\lin{x}{e_0}{e_1}\right) \cdot e_2
                          &\rewritesto  \lin{x}{e_0}{e_1 \cdot e_2}           \\
f\left(\lin{x}{e_0}{e_1}, \hdots \right)
                          &\rewritesto  \lin{x}{e_0}{f(e_1, \hdots)}          \\
(\lin{x}{e_0}{e_1}, e_2)  &\rewritesto  \lin{x}{e_0}{(e_1,e_2)}               \\
\pi_0(\lin{x}{e_0}{e_1})  &\rewritesto  \lin{x}{e_0}{\pi_0 e_1}               \\
\gen_{i=0}^n(\lin{x}{e_0}{e_1})
                          &\rewritesto  \lin{x}{\left(\gen_{i=0}^n e_0\right)}
                                               {\gen_{i=0}^n \msub{x\mapsto x[i]} e_1} \\
\ssum_{i=0}^n(\lin{x}{e_0}{e_1})
                          &\rewritesto  \lin{x}{\left(\gen_{i=0}^n e_0\right)}
                                               {\ssum_{i=0}^n \msub{x\mapsto x[i]} e_1} \\
(\lin{x}{e_0}{e_1})[a]    &\rewritesto  \lin{x}{e_0}{e_1[a]}                  \\
\label{rule:let-lift-indicator}
\ind{p}\cdot(\lin{x}{e_0}{e_1})
                          &\rewritesto  \lin{x}{\ind{p}\cdot e_0}{\ind{p}\cdot e_1} \\
\lin{x}{\left(
    \lin{y}{e_0}{e_1}
        \right)}{e_2}     &\rewritesto  \lin{x}{e_1}{\lin{y}{e_0}{e_2}}       %
\end{align}
\end{framed}
\caption{Let-Lifting}
\label{fig:let-lift}
\endgroup
\end{figure}

For instance, consider the Let-Lift normalization pass (Figure \ref{fig:let-lift}).  The goal of this pass is to move all $\kwd{let}$ bindings to the outermost level of the expression, placing the expression into a statement-block/return-value form.  We make no assumptions about the normality of the input expression in order to achieve this.  We can state the normal form precisely, by introducing a new grammatical category $Le$ and reinterpreting $e$ in this context to lack the $\kwd{let}$ form.

$$ Le ::= e~|~\kwd{let} x \keq e \kwd{in} Le $$

In order to see why this normal form must be achieved, consider some expression $e$ which has been fully normalized using the let-lifting rules.  Now suppose for the sake of contradiction that it does not have the specified syntactic normal form.  Then there must be some expression context $C$ such that $e = C[\kwd{let} x \keq e_1 \kwd{in} e_2]$; In particular, there exists some expression immediately wrapped around the let-binding.  By exhaustive case analysis, this must match one of the left-hand-side expressions in the let-lifting rewrite rules.  However, this is impossible if $e$ was fully normalized.  Therefore, our supposition was faulty.

To see why the let-lifting rules are correct, we can make a simple inductive argument over the rewrite rules as cases.  For each rule, we can apply evaluation to the left and right hand sides.  For instance, applying to the product case of the left-lift rules, we get

\begin{align*}
\EvalExpr{(\kwd{let} x \keq e_0 \kwd{in} e_1)\cdot e_2\,}{\,\sigma}
&= \EvalExpr{\kwd{let} x \keq e_0 \kwd{in} e_1\,}{\,\sigma} \cdot \EvalExpr{e_2}{\sigma} & \textrm{by 4.4}\\
&= \EvalExpr{e_1\,}{\,\sigma\msub{x\mapsto \EvalExpr{e_0\,}{\sigma}}} \cdot \EvalExpr{e_2}{\sigma}  & \textrm{by 4.9} \\
&\hspace{0.5in}\textit{(since $x\not\in FV(e_2)$ $\hdots$)}\\
&= \EvalExpr{e_1\,}{\,\sigma\msub{x\mapsto \EvalExpr{e_0}{\sigma}}} \cdot
   \EvalExpr{e_2\,}{\,\sigma\msub{x\mapsto \EvalExpr{e_0}{\sigma}}}  \\
&= \EvalExpr{e_1\cdot e_2\,}{\,\sigma\msub{x\mapsto \EvalExpr{e_0}{\sigma}}}  & \textrm{by 4.4} \\
&= \EvalExpr{\kwd{let} x \keq e_0 \kwd{in} e_1 \cdot e_2\,}{\,\sigma}  & \textrm{by 4.9}\\
\end{align*}

We will not continue making correctness arguments at this level of detail, but it is instructive to see that doing so (at a level amenable to mechanization) is possible.

Using a similar process, we can argue that the let-lifting pass is cost neutral because each constituent rule is.  Examining that same case,

\begin{align*}
\CostExpr{(\kwd{let} x \keq e_0 \kwd{in} e_1)\cdot e_2\,}{\,\sigma}
&= 1 + \CostExpr{\kwd{let} x \keq e_0 \kwd{in} e_1\,}{\,\sigma} + \CostExpr{e_2}{\sigma} & \textrm{by 4.25}\\
&= 1 + \left(\CostExpr{e_0}{\sigma} + \CostExpr{e_1}{\sigma}\right) + \CostExpr{e_2}{\sigma} & \textrm{by 4.32}\\
&= \CostExpr{e_0}{\sigma} + \left(1 + \CostExpr{e_1}{\sigma} + \CostExpr{e_2}{\sigma}\right) \\
&= \CostExpr{e_0}{\sigma} + \CostExpr{e_1 \cdot e_2}{\sigma} & \textrm{by 4.25}\\
&= \CostExpr{\kwd{let} x \keq e_0 \kwd{in} e_1 \cdot e_2\,}{\,\sigma} & \textrm{by 4.32}\\
\end{align*}

We will run into one common technical obstacle in the structure of cost-neutrality proofs.  Because the cost-model makes a special exception for terms of the form $\ssum\ind{p}\cdots$, our inductive arguments will have to distinguish sub-cases of $\ssum e$ depending on whether $e = \ind{p}\cdot e'$ or not.  When dealing with the $\ssum\ind{p}\cdots$ sub-case, we may simply assume that the two relevant rewrite rules are applied back-to-back.  This introduces some technical issues into the structure of induction, but not in ways that are illuminating enough to dwell on.

Now that we have the general structure of the argument out of the way, we can focus on those odd or exceptional cases.  These are the rules for $\gen$, $\ssum$, and $\ind{p}$.  If we first focus on rule \ref{rule:let-lift-indicator} for predicates, notice that it would also be correct to \emph{not} duplicate $\ind{p}$ as being wrapped around $e_0$.  However, (i) doing so is safe and (ii) doing so is necessary to ensure cost-neutrality under our given cost model.

The other major non-obvious rules are for $\gen$ and $\ssum$.  These can be viewed one of two ways.  If on the one hand we view $\gen$ as like a $\lambda$-abstraction, and $\kwd{let}$ as sugar for a red-ex, then this rule is an instance of let-lifting or lambda-lifting~\cite{Johnsson:1985:LLT} transformations that lift all new function declarations to a top level.  In the process of doing so, some free variables in the expression $e_0$ must be turned into explicit parameters (i.e. parameterization over $i$).  On the other hand, if we view $\gen$ or $\sum$ as more like a loop, then we can view these transformations as instances of loop-fission.  Regardless we can verify equivalence algebraically.

Note that our rules for $\gen$ and $\ssum$ are cost-neutral because we chose to discount the cost of memory traffic, reading and writing.  This inaccuracy is acceptable because the purpose of our cost model is to reproduce Griewank's arguments for bounding the number of operations on continuous values.

\subsection{Pair Elimination}

Before we discuss the pair elimination pass, we must define a type isomorphism corresponding to the difference between struct-of-arrays and array-of-structs data structures.  This isomorphism is 
$$ [n](T_0\times T_1) \cong [n]T_0 \times [n]T_1 $$

If a type has array types fully distributed over pair types (i.e. the type can no longer be rewritten in the rightward direction) then we say the type is in \emph{struct-of-arrays} form.  The goal of pair elimination is to place all intermediate values into struct-of-arrays form and then to destructure those intermediaries.  The only pair constructs remaining in the program are projections of inputs, and pair constructions of final outputs.

\begin{figure}[t]
\begingroup
\newcommand{\lin}[3]{\kwd{let} #1 \keq #2 \kwd{in} #3}
\arraycolsep=1.8pt
\begin{framed}
\begin{align}
\pi_0 (e_0, e_1)          &\rewritesto  e_0                                   \\
\pi_0 \left(\gen_{i=0}^n e\right)
                          &\rewritesto  \gen_{i=0}^n \pi_0 e_0 \\
\gen_{i=0}^n (e_0, e_1)
                          &\rewritesto  \left( \gen_{i=0}^n e_0,
                                               \gen_{i=0}^n e_1 \right) \\
\pi_0 \left(e[a]\right)   &\rewritesto  (\pi_0 e)[a] \\
(e_0,e_1)[a]              &\rewritesto  (e_0[a],e_1[a]) \\
\pi_0 \ind{p}\cdot e      &\rewritesto  \ind{p}\cdot (\pi_0 e) \\
\ind{p}\cdot (e_0,e_1)    &\rewritesto  \left( \ind{p}\cdot e_0,
                                               \ind{p}\cdot e_1 \right) \\
\lin{x : [n_0]\cdots[n_k](T_0 \times T_1)}{e_0}{e_1} &\rewritesto
    \lin{x_0 : [n_0]\cdots[n_k]T_0}{\pi_0 e_0}{} \\
&\phantom{\rewritesto}\,
    \lin{x_1 : [n_0]\cdots[n_k]T_1}{\pi_1 e_0}{} \\
&\phantom{\rewritesto}\,
    \msub{x \mapsto (x_0,x_1)} e_1       %
\end{align}
\end{framed}
\caption{Pair-Elimination}
\label{fig:pair-elim}
\endgroup
\end{figure}

Pair elimination (Figure \ref{fig:pair-elim}) assumes that terms are already in let-lifted normal form.  Additionally, and non-trivially, pair-elimination assumes that all input (free-variable) types and the output type are in struct-of-array form\footnote{This constraint can be relaxed, but at the expense of handling a lot of uninsightful details.}.  The resulting normal form can be captured grammatically as
\begin{align*}
Le  ::= &~ \kwd{let} x \keq e \kwd{in} Le ~|~ Oe\\
Oe  ::= &~ (Oe,Oe) ~|~ e \\
Ie  ::= &~ \pi_0~Ie ~|~ \pi_1~Ie ~|~ x \\
\end{align*}
where the grammatical class $e$ in this context lacks $\kwd{let}$, pair-construction, and projection, but includes $Ie$, input expressions.  All variables $x$ occurring in an input expression must be free at the top-level.

The argument for why the pair-elimination rules achieve this normal form is more complicated.  First, note that the rule for $\kwd{let}$ bindings will de-structure all intermediaries with pair types until all intermediary variables have either scalar or tensor types.  Free variables (inputs) may have struct-of-array type.

Observe that the pair-elimination rules are not guaranteed to preserve typing (e.g. interchange of projection and tensor access) but do preserve typing up to isomorphism.  On this basis of this invariant, we can make a type-based argument that the normal form is achieved.  First, suppose that after running to termination, there remains some projection $\pi_k e$ where $e$ is not a free (input) variable $x$.  If $e$ is an intermediary variable, this violates our typing invariant.  Then consider all other possible grammatical forms of $e$.  If $e$ is a scalar expression (constant, plus, times, summation, or black-box function) then the typing invariant must have been violated.  Otherwise, we can identify a rewrite rule which hasn't been applied.  Likewise, suppose pair-construction occurs somewhere other than in the final, output position.  Then, the type invariant has been violated, or some rewrite rule can be applied.

The correctness of pair-elimination is complicated by the introduction of non-sensical intermediary expressions.  For instance,
$$\kwd{let} x : [n](\R \times \R) \keq \gen_{i=0}^n (x[i],x[i]) \kwd{in} \hdots
  \rewritesto\kwd{let} x_0 : [n]\R \keq \pi_0\gen_{i=0}^n (x[i],x[i]) \kwd{in} \hdots$$
we can recover a consistent meaning for such terms by conjugating operations with the type isomorphism.  For instance, when we have $\pi_0 e$ with $e : [n](T_0,T_1)$,
then let $\pi_0 e \cong \gen_{i=0}^n\pi_0 e[i]$.  By providing this sort of meaning to every otherwise non-sensical expression, all of the pair-elimination rules can then be verified in the usual way.

A slight remaining concern might be that some of these non-sensical terms will remain in the expression after the completion of the pair-elimination rewrites.  However, this cannot happen by the preceding normalization argument.

Cost-neutrality is trivial to verify for every rule except for the decomposition of $\kwd{let}$ expressions, which results in a duplicated right-hand-side.  To argue that this (surprisingly) doesn't introduce any additional cost, we must argue that after the projections are fully pushed-in, no costly terms remain shared between the two new right-hand-sides.  To argue this, recall in our earlier argument that pair-elimination does not interact with scalar operations ($+$, $\cdot$, $f(\hdots)$, $\ssum$).  However, non-scalar operations are cost-neutral in our model.  Therefore any duplication of non-scalar operations is allowable.  As projections get pushed down the right-hand-sides, they will pass through (and therefore commit to duplicating) these cost-neutral operations.  However, due to the type-isomorphism and normal form argument, the projections must either end up on an input variable (in which case the entire right-hand-side had no cost) or end up interacting with a pair-constructor.  If the projections encounter a pair-constructor, then one right-hand-side will reduce to have only the first half of that constructor and the other right-hand-side will reduce to have the second half.  At this point, there is no duplication, and so cost-neutrality has been achieved.

\subsection{Gen Pushout}

\begin{figure}[t]
\begingroup
\newcommand{\lin}[3]{\kwd{let} #1 \keq #2 \kwd{in} #3}
\arraycolsep=1.8pt
\begin{framed}
\begin{align}
\left(\gen_{i=0}^n e\right)[a]  &\rewritesto \msub{i \mapsto a} e           \\
\left(\ind{p}\cdot e\right)[a]  &\rewritesto \ind{p}\cdot e[a]              \\
\ind{p}\cdot\left(\gen_{i=0}^n e\right)
                                &\rewritesto \gen_{i=0}^n \ind{p}\cdot e    \\
\ind{p_0}\cdot\ind{p_1}\cdot e  &\rewritesto \ind{p_0\conj p_1}\cdot e      %
\end{align}
\end{framed}
\caption{Gen-Pushout}
\label{fig:gen-pushout}
\endgroup
\end{figure}

The third pass, Gen-Pushout (Figure \ref{fig:gen-pushout}) is very simple.  It is designed to be applied after pair-elimination and further normalizes the program to ensure that all Gen operators $\gen$ occur at the outer-most position of right-hand-side expressions.  This normal form is given by the following grammar.
\begin{align*}
Le  ::= &~ \kwd{let} x \keq Ge \kwd{in} Le ~|~ Oe\\
Ge  ::= &~ \gen_{i=0}^n Ge ~|~ e \\
Oe  ::= &~ (Oe,Oe) ~|~ GOe \\
GOe ::= &~ \gen_{i=0}^n GOe ~|~ e \\
Ae  ::= &~ Ae[a] ~|~ x \\
Ie  ::= &~ Ie[a] ~|~ PIe \\
PIe ::= &~ \pi_0~PIe ~|~ \pi_1~PIe ~|~ x \\
\end{align*}
where the remaining grammatical class $e$ now can include terms from $Ie$ and from $Ae$, but excludes cases for tensor generation or indexing.  The class $e$ now consists solely of indicator functions, big summations, and other scalar operations.

A simple case analysis argument will prove that this normal form is achieved.  For correctness, the argument is likewise a simple verification of these four rules.  Finally, cost-neutrality is trivial, with the observation that the first rule may actually decrease costs.

\subsection{Single-Static-Assignment for Tensor Code}

Once tensor code is flattened by the three preceding passes, we can perform a final normalization into our single-static-assignment form (Figure \ref{fig:ssa-grammar}).  Doing this has a variety of benefits.  Because we are working in a purely functional language, we can perform structural memo-ization of right-hand-side expressions in order to achieve common sub-expression elimination.   Likewise, we can accomplish other optimizations that rely on approximating program equality such as $e + e = 2\cdot e$.  Most importantly, we will be able to argue that performing adjoint-differentiation on this SSA form introduces no more than a constant factor overhead---while remaining purely functional.

\begin{figure}[t]
\begingroup
\newcommand{\h}{\harpoon}
\begin{tabular}{rcll}
\toprule
$S$  & $::=$ & $\kwd{let} X \keq E \kwd{in} S$      & statement binding \\
     &  $|$  & $X$                                  & lone output value \\
     &  $|$  & $(X_0,\hdots,X_k)$                   & output tuple \\
\midrule
$E$  & $::=$ & $\displaystyle\pi_{k_0}\cdots\pi_{k_l} x$
             & input variable \\
     &  $|$  & $\displaystyle\gen_{\h\imath} c$
             & constant number \\
     &  $|$  & $\displaystyle\gen_{\h\imath} \left( \ind{p_0}\cdot X_0[\h\imath] +
                                                    \ind{p_1}\cdot X_1[\h\imath] \right)$
             & binary addition \\
     &  $|$  & $\displaystyle\gen_{\h\imath} \ind{p}\cdot f\left(X_0[\h\imath], \hdots\right)$
             & map scalar function \\
     &  $|$  & $\displaystyle \gen_{\h g} \ssum_{\h\imath_0, \h\imath_1, \h s}
                \ind{p} \cdot X_0[\h\imath_0] \cdot X_1[\h\imath_1]$
             & binary tensor contraction product\\
     &  $|$  & $\displaystyle \gen_{\h g} \ssum_{\h\imath, \h s}
                \ind{p} \cdot X_0[\h\imath]$
             & unary tensor contraction\\
\midrule
     &       & \textit{short-hand notations} \\
     &       & C & constant number \\
     &       & $\displaystyle \left( \ind{p_0}\cdot X_0 + \ind{p_1}\cdot X_1 \right)$
             & simple addition \\
     &       & $\displaystyle \ind{p}\cdot f\left(X_0, \hdots\right)$
             & map scalar function \\
\midrule
$X$         & &                                     & SSA intermediate variable \\
$\h\imath$  & &                                     & list of index variables \\
\bottomrule
\end{tabular}
\endgroup
\caption{Grammar for Tensor SSA Normal Form}
\label{fig:ssa-grammar}
\end{figure}

The SSA form is defined grammatically (Figure \ref{fig:ssa-grammar}) with reference to the original language semantics.  These SSA forms can become verbose and complicated by keeping track of index variables.  In the case of constants, simple additions, and scalar functions, the indexing is trivial and can be omitted (see short-hand notations).  In the case of the tensor contraction product, this is not true.  Therefore to simplify our notation in this paper, we adopt the convention of an over-arrow (like a vector) to designate a list of index variables and ranges.  That is,
$$ \harpoon\imath = \{i_0\in[0,n_0), \hdots i_k\in[0,n_k)\} $$
As a further simplification, and because the lexical scope of each index variable is now much more limited, we can re-name all index variables to canonical names based on the order in which they are introduced.  This allows for the short-hand notation $\ind{p}\cdot(X_0 + X_1)$ to be meaningful, because all of the index variables referred to in $p$ can be assumed to use these canonical names.

A few important algebraic observations can be made about this SSA form.  First, it resembles representations for polynomials with two fundamental operators: addition and multiplication.  In the degenerate case\footnote{coincidentally, de-generate happens to mean literally without gen in our context, i.e. without $\gen$} where $\harpoon\imath = \{\}$, this is just scalar polynomials.  However, in the tensorial general case, multiplication has become much more complicated, accounting for all kinds of intermixing of products, transpositions, and sparsity structures.  Second, (and relatedly) the ``structure'' and ``sparsity'' of any given tensor-contraction is defined by the predication $\ind{p}$.  For instance, consider the product $\gen_{g_0,g_1}\ssum_{i_0,j_0,i_1,j_1} \ind{p}\cdot A[i_0,j_0]\cdot B[i_1,j_1]$.  If $p$ is $(g_0\keq i_0\keq i_1 \conj g_1\keq j_0\keq j_1)$, then the product is a point-wise multiplication running in quadratic time, but if $p$ is $(g_0\keq i_0 \conj j_0\keq i_1 \conj j_1\keq g_1)$, then the product is a matrix-matrix multiplication running in cubic time.

We adopt the convention that different $X$ variables are identical or not depending on syntactic equality of their right-hand-side definitions.  For this reason, it is helpful to maintain certain ordering conventions, making use of some arbitrary total ordering imposed on $X$ names:
\begin{itemize}
    \item the terms in a simple addition are always in-order.
    \item the factors in a contraction product are always in-order,
          and indexing variables are renamed appropriately.
    \item the index variables are always named canonically based on their order of occurrence.
\end{itemize}

The preceding defines the post-condition normal form characterizing our SSA for tensors.

\begin{figure}[t]
\begingroup
\newcommand{\D}[1]{\SSABracket{#1}}
\newcommand{\lin}[3]{\kwd{let} #1 \keq #2 \kwd{in} #3}
\newcommand{\h}{\harpoon}
\newcommand{\hi}{\h\imath}
\arraycolsep=1.8pt
\begin{framed}
\begin{align}
\D{\lin{x}{e_0}{e_1}}       &\rewritesto \lin{X}{\D{e_0}}{\D{\msub{x\mapsto X}e_1}} \\
\D{X}                       &\rewritesto X \\
\D{\gen_{\hi}\ind{p}\cdot\gen_{j=0}^n\cdot e} &\rewritesto \D{\gen_{\hi,j}\ind{p}\cdot e} \\
\D{\gen_{\hi}\ind{p_0}\cdot\ind{p_1}\cdot e} &\rewritesto \D{\gen_{\hi}\ind{p_0\conj p_1}\cdot e} \\
\D{\gen_{\hi}\ind{p}\cdot c}            &\rewritesto C \\
\begin{split}
\D{\gen_{\h g}\ind{p}\cdot \left(\pi_{j_0}\cdots\pi_{j_l} x\right)[a_0,\hdots,a_k]}
                    &\rewritesto   \lin{X}{\pi_{k_0}\cdots\pi_{k_l} x}{} \\
         &\phantom{\rewritesto}\ \ \gen_{\h g} \ssum_{i_0,\hdots,i_k}
                                    \ind{p\conj\begin{array}{l}
                                        i_0\keq a_0 \conj\\
                                        \cdots \conj\\
                                        i_k\keq a_k
                                    \end{array}}\cdot
                                    X[i_0,\hdots,i_k]
\end{split}
\end{align}
\end{framed}
\caption{Converting Flattened Tensor Code into SSA (first half)}
\label{fig:flat-to-ssa-1}
\endgroup
\end{figure}


\begin{figure}[t]
\begingroup
\newcommand{\D}[1]{\SSABracket{#1}}
\newcommand{\lin}[3]{\kwd{let} #1 \keq #2 \kwd{in} #3}
\newcommand{\h}{\harpoon}
\newcommand{\hi}{\h\imath}
\arraycolsep=1.8pt
\begin{framed}
\begin{align}
\begin{split}
\D{\gen_{\hi}\ind{p}\cdot(\ind{p_0}\cdot e_0 + \ind{p_1}\cdot e_1)}
                    &\rewritesto   \lin{X_0}{\D{\gen_{\hi}\ind{p\conj p_0}\cdot e_0}}{} \\
         &\phantom{\rewritesto}\ \ \lin{X_1}{\D{\gen_{\hi}\ind{p\conj p_1}\cdot e_1}}{} \\
         &\phantom{\rewritesto}\ \ \left( \ind{p\conj p_0}\cdot X_0 + \ind{p\conj p_1}\cdot X_1 \right)
\end{split}\\
\begin{split}
\D{\gen_{\hi}\ind{p}\cdot e_0\cdot e_1}
                    &\rewritesto   \lin{X_0}{\D{\gen_{\hi}\ind{p}\cdot e_0}}{} \\
         &\phantom{\rewritesto}\ \ \lin{X_1}{\D{\gen_{\hi}\ind{p}\cdot e_1}}{} \\
         &\phantom{\rewritesto}\ \ \gen_{\h g}\ssum_{\h\imath_0 \h\imath_1}
                                    \ind{p~\conj \h g \keq \h\imath_0 \keq \h\imath_1}\cdot
                                    X_0[\h\imath_0] \cdot X_1[\h\imath_1]
\end{split} \\
\begin{split}
\D{\gen_{\hi}\ind{p}\cdot f(e_0,\hdots)}
                    &\rewritesto   \lin{X_0}{\D{\gen_{\hi}\ind{p}\cdot e_0}}{}
                               \ \ \cdots\\
         &\phantom{\rewritesto}\ \ \ind{p}\cdot f(X_0,\hdots)
\end{split}\\
                                    %
\begin{split}
\D{\gen_{\hi}\ind{p}\cdot\ssum_{j=0}^n e}
                    &\rewritesto \lin{X}{\D{\gen_{\hi,j} \ind{p}\cdot e}}{}\\
         &\phantom{\rewritesto}\ \ \gen_{\h g}\ssum_{\h i,j} \ind{p~\conj \h g \keq \hi}\cdot X[\hi,j] %
\end{split}
\end{align}
\end{framed}
\caption{Converting Flattened Tensor Code into SSA (second half)}
\label{fig:flat-to-ssa-2}
\endgroup
\end{figure}

We can convert from flattened Tensor code into the SSA Normal form via a transformation $\SSABracket{e}$ (Figures \ref{fig:flat-to-ssa-1} \& \ref{fig:flat-to-ssa-2}).  The rules detailed in the figure accomplish the spirit of the transformation but require some additional fix-ups.  To fix those, note the following.  (1) Extraneous intermediary variables may be introduced, resulting in $\kwd{let} X_0 \keq X_1 \kwd{in} \hdots$, which may be removed via a simple substitution.  (2) The transform produces an unbalanced sequence of $\kwd{let}$-bindings, which can be trivially flattened into a single sequential block via right-association.  (3) the output position may have a primitive right-hand-side expression rather than a single variable; which may be resolved by naming the expression with a new let-binding.  (4) common sub-expression-elimination, and normalization of ordering may be fixed as detailed above.  In our compiler prototype, the entire representation is instead stored as a DAG via memoization and pointer aliasing; which incidentally handles many of these details.

Lastly, observe that most rules process expressions of the form $\gen_{\harpoon\imath}\ind{p}\cdot e$.  In the degenerate cases $p = \kwd{true}$ and $\harpoon\imath = \{\}$, but in general this can be viewed as the propagation of a computation mask over all primitive operations.  As we will see shortly, this propagation of predicates is essential for maintaining cost.

First, we would like to argue that this pass produces the desired normal form.  Taking into account the preceding finer details, we can see that the result of rewriting ensures that every right-hand-side expression conforms to one of the normal form cases, and that all intermediaries are assigned to new variables.

The argument for correctness can be made via a simple structural recursion built on top of a case-by-case analysis.  The one potentially trouble-some case is black-box functions, where the introduction of predicates to the arguments is not justified by a common algebraic law.  In this case, the fact that the result of the function is ignored anyway allows us to modify the inputs arbitrarily in service of short-circuiting the computation of arguments as well.

Finally, and least obviously, we claim that the conversion to SSA is cost-neutral, despite what appears to be very gratuitous introduction of intermediary variables and loops.  Before digging into a formal argument, it is worth re-emphasizing that our cost model is designed to count only arithmetic operations performed on non-indexing values.  As suggested earlier, if each line of this SSA was executed as a predicated instruction on a GPU or other wide SIMD machine, then the real cost would be substantially different.  Because our cost model is only built to allow us to bound the cost of differentiation in the style of Griewank, this choice is necessary and appropriate.

Our formal argument will proceed in the usual way, by arguing that each individual rule is cost-preserving.  The first five rules are trivially so.  When we get to the case of indexing arithmetic, we introduce a big summation on the right-hand-side, which has the potential to increase the cost of the left-hand-side (which is $0$).  However, by construction, for each setting of the index variables $\harpoon g$, there is at most $1$ valid assignment of $i_0, \hdots, i_k$.  In this case of summing over one or zero items, the big summation introduces no additional cost.  This same line of argument applies to the final rule for converting big summations, and to the rule for converting products.  However, in those cases (and in all other remaining cases) we must also account for the recursive cost of sub-expressions which have been lifted up into intermediary variables.  In all these cases observe that the generator-predicate form resolves to a cost-multiplier that reflects the number of in-range index variable assignments satisfying $p$.  In our algebraic proof, this manifests as an application of distributivity between the loop multiplier and the summation of in-loop work.

\section{Derivatives}
\label{sec:derivatives}

\subsection{What is the derivative of an expression?}
Recall (\S\ref{sec:expressions-are-functions}) that given types for all free-variables of an expression $e$, we may interpret $e$ as denoting a smooth, total function between finite dimensional vector spaces $f : \R^{n_I} \to \R^{n_O}$.  Therefore, in the same way that we can talk about the derivative of a function (with respect to one or more of its arguments) we can talk about the derivative of an expression (with respect to one or more of its free variables).

The \emph{Total Derivative} or \emph{Forward-Mode} derivative of $f : A \to B$ is the unique function $Df : A \to A \to B$ s.t. $Df$ is linear in its second argument, and s.t. $Df$ is the ``closest approximation'' satisfying
$$ f(x + dx) \approx f(x) + Df(x,dx) $$
Under suitable differentiability assumptions, this definition coincides with a coordinate-wise definition in terms of partial derivatives:
$$ Df(x,dx)_j = \sum_{i=0}^{n_I} \left.\frac{\partial f_j}{\partial x_i}\right|_{x} \cdot dx_i $$
where $\left.\frac{\partial f_j}{\partial x_i}\right|_{x}$ is the evaluation of the partial at base-point $x$.  Note that we may also compute any partial from the total derivative as $\left.\frac{\partial f_j}{\partial x_i}\right|_{x} = Df\left(x,\gen_k^{n_I}\ind{k=i}\right)$, or in the particular case of a scalar function $f : \R \to \R$, we may reproduce the most familiar definition of a function's derivative: $\frac{df}{dx} = Df(x,1)$.

The \emph{Adjoint Derivative} or \emph{Reverse-Mode} derivative of $f : A \to B$ is the function $D^Tf : A \to B \to A$ with $D^Tf(x)$ linear, s.t. $D^Tf(x) = (Df(x))^T$ when we view those linear functions as matrices.  While this can be defined in terms of coordinates, it can also be defined according to the algebraic universal property of transposition:
$$ \forall dx,dy: \inner{dy}{Df(x,dx)} = \inner{D^T(x,dy)}{dx} $$
where $\inner{x_0}{x_1}$ is an assumed standard inner product on the vector space.

Using these two primitives, we can define the gradient of a function $f : A \to \R$ as $\nabla_x f = D^Tf(x,1)$ or the Hessian-vector-product at a point $x$ as $(H_x f) dx = D(\nabla_x f)(x,dx)$.

In this formulation, for $f: A \to B$ and $g : B \to C$, the chain rule is realized as
$$ D(f\circ g)(x,dx) = Df( g(x), Dg(x, dx) ) $$

\subsection{The Total (forward) Derivative}

\begin{figure}[t]
\begingroup
\newcommand{\D}{\deriv}
\arraycolsep=1.8pt
\centering
\begin{framed}
\begin{align}
\D{x}{\sigma\msub{x\mapsto dx}} &\rewritesto dx \\
\D{x}{\sigma} &\rewritesto 0,~~~~\textrm{where } x\not\in \sigma \\
\D{c}{\sigma} &\rewritesto 0 \\
\D{e_0 + e_1}{\sigma} &\rewritesto \D{e_0}{\sigma} + \D{e_1}{\sigma} \\
\label{rule:fwd-leibniz-prod-rule}
\D{e_0 \cdot e_1}{\sigma} &\rewritesto \D{e_0}{\sigma}\cdot e_1 + e_0\cdot\D{e_1}{\sigma} \\
\D{f\left( e_0,\hdots \right)}{\sigma} &\rewritesto Df\left( e_0,\hdots;\D{e_0}{\sigma},\hdots \right) \\
\D{\left( e_0,e_1 \right)}{\sigma} &\rewritesto \left( \D{e_0}{\sigma},\D{e_1}{\sigma} \right) \\
\D{\pi_0~e}{\sigma} &\rewritesto \pi_0~\D{e}{\sigma} \\
\D{\gen_{i:n} e}{\sigma} &\rewritesto \gen_{i:n} \D{e}{\sigma} \\
\D{\ssum_{i:n} e}{\sigma} &\rewritesto \ssum_{i:n} \D{e}{\sigma} \\
\D{e[a]}{\sigma} &\rewritesto \D{e}{\sigma}[a] \\
\D{\ind{p}\cdot e}{\sigma} &\rewritesto \ind{p}\cdot\D{e}{\sigma} \\
\D{\kwd{let} x \keq e_0 \kwd{in} e_1}{\sigma} &\rewritesto
  \begin{array}[t]{rrcl}
    \kwd{let}& x  &\keq& e_0 \kwd{in} \\
    \kwd{let}& dx &\keq& \D{e_0}{\sigma} \kwd{in} \\
    & \multicolumn{3}{l}{\D{e_1}{\sigma\msub{x \mapsto dx}}} \\
  \end{array}
\end{align}
\end{framed}
\caption{Total (Forward) Derivative for Core Tensor Language}
\label{fig:atl_derivative}
\endgroup
\end{figure}

Given an environment $\sigma$ mapping input variables $x$ to new corresponding differential variables $dx$, we define the total derivative (Figure \ref{fig:atl_derivative}) as a meta-linguistic source-to-source transform on well-formed expressions $e$, s.t. if $e$ corresponds to $f$, then $\deriv{e}{\sigma}$ corresponds to $Df$.  (Note that we may control which variables are differentiated or not by controlling the contents of $\sigma$.)

The correctness of this transformation (that it really represents the total derivative) follows from commonly known properties of derivation (e.g. Leibniz multiplication rule, linearity of the derivative).  Maybe the only exception is the rule for $\kwd{let}$ expressions, which is (non-obviously) a consequence of the chain rule.  To see why, recall that $\kwd{let}$ expressions may be viewed as sugar for beta-reductions in the lambda-calculus.  Under this interpretation, the expression $\kwd{let} x \keq e_0 \kwd{in} e_1$ is $(\lambda x. e_1)(e_0)$.  Let $y$ range over the variables differentiated by $\sigma$ so that we may interpret $e_0$ as a function $f_0(y)$ and $e_1$ as a function $f_1(x,y)$.  Then the $\kwd{let}$ expression is the composition $f_1( f_0(y), y )$, whose derivative (according to the chain rule) is $Df_1( f_0(y), y, Df_0(y,dy), dy )$.  Letting $x = f_0(y)$ and $dx = Df_0(y,dy)$ and converting back into $\kwd{let}$ expression form, this is just the expression on the right of the rewrite.

It is a well-known fact that without normalization of some form, this transform may cause combinatorially large cost-growth.  For instance, a simple scalar product of $k$ variables is represented by an expression of size $k$, while its derivative according to the preceding rules will have size $O(k^2)$.  If intermediaries in the scalar product are bound via $\kwd{let}$ expressions, then this is no longer true.  The derivative expression has size $O(k)$ as well\footnote{This example is often given as evidence of a short-coming of ``symbolic differentiation'' relative to ``automatic differentiation''.  Rather than dwell on various coincidental features of those terminologies, we believe this demonstration gets at the real heart of the matter.  Namely, sharing, as enabled via DAGs or named intermediate expressions is crucial.  The shortcoming of most ``symbolic'' approaches is simply an over-commitment to maintaining all expressions as trees without $\kwd{let}$ bindings}.

\begin{theorem}[cost-efficiency of Total Derivative]
\label{thm:fwd-deriv-cost}
Let $e$ be a well-typed expression in Tensor SSA form.  Then, $\CostExpr{\deriv{e}{\sigma}}{} \leq 4\cdot\CostExpr{e}{}$.
\end{theorem}
\begin{proof}
(sketch of proof) By case analysis of the SSA normal form (Figure \ref{fig:ssa-grammar}).  Exactly twice as many $\kwd{let}$ expressions are introduced, and the right-hand sides of the new differential $\kwd{let}$ expressions cost at most $3$ times (Leibniz product rule \ref{rule:fwd-leibniz-prod-rule}) as much as the original right-hand-side, under the standard work-cost model.  Therefore, the derivative cost is at most $4\times$ the original cost. ($(3\times) + (1\times)$)
\end{proof}

\subsection{The Adjoint (reverse) Derivative}

The adjoint derivative of an expression $\derivT{e}{\sigma}$ is significantly more complicated to specify programmatically than the total derivative.  Historically, this has led to a number of conceptual and mnemonic devices for understanding it.  The earliest such device is the Wengert Tape, which is traditionally understood via a formulation purely in terms of partial derivatives.  Derivatives are accumulated onto the tape as the execution proceeds until a final point is reached and the tape is unwound to propagate differentials backwards.  Another approach represents the linear part of the derivative as a DAG, with multiplication living on the edges of the graph and fan-in to nodes representing addition (of scalar values).  In this conception, transposition simply consists of reversing the direction of all the arrows.  Finally, Conal Elliott\cite{Elliott:2018:SEA} has recently advocated for handling transposition via the duality of biproduct categories.

We rely on a different approach, leveraging the (defining) universal property of transposition.  Re-stated in terms of our expressions, let $e$ be an expression with free variable(s) of type $x : A$, and named output of type $y : B$.  (If $e$ has more than one free variable, assume they have been grouped into nested tuples without loss of generality)  Let $f:A\to B$ be the function corresponding to $e$.  Then $e^T$ is an expression representing $f^T$ if and only if
$$ \forall x'\in A,y'\in B: \inner{y'}{\msub{x\mapsto x'}e}_B = \inner{\msub{y\mapsto y'}e^T}{x'}_A $$
This formulation immediately suggests that if we have some way to algebraically rewrite the left-hand-side of this equation into the right-hand-side, then we will have a correct by construction method for transposing linear expressions in our language.

\begin{figure}[t]
\begingroup
\newcommand{\I}{\inner}
\newcommand{\+}{\mplus}
\newcommand{\0}{\mzero}
\newcommand{\E}[1]{%
  \color{meta}\left\llbracket
    \normalcolor #1
  \color{meta}\right\rrbracket
    \normalcolor
}
\arraycolsep=1.8pt
\begin{framed}
Grammar Extension
\begin{align*}
e^L  ::= &~\mwd{let} dx \meq \deriv{e}{\sigma} \mwd{in} e^L ~|~ \kwd{let} x \keq e \kwd{in} e^L ~|~ e^S \\
e^S  ::= &~\ e^I \+ e^S ~|~ e^I \\
e^I  ::= &~ \I{e}{e} ~|~ \I{e}{e}_{\!\msub{p}} ~|~ \0 \\
\end{align*}
Semantics via de-sugaring
\begin{align*}
\E{\mwd{let} dx \meq \deriv{e}{\sigma} \mwd{in} e^L}&\to \kwd{let} dx \keq \deriv{e}{\sigma} \kwd{in} \E{e^L} \\
\E{e^I \+ e^S}                      &\to \E{e^I} + \E{e^S} \\
\E{\I{e_0}{e_1}}                    &\to e_0 \cdot e_1                  &(e_0,e_1 : \R) \\
\E{\I{e_0}{e_1}}                    &\to \E{\I{\pi_0~e_0}{\pi_0~e_1}} +
                                         \E{\I{\pi_1~e_0}{\pi_1~e_1}}   &(e_0,e_1:(T_0,T_1)) \\
\E{\I{e_0}{e_1}_{\!\msub{p}}}       &\to \ssum_{\harpoon\imath}^{\harpoon n} \ind{p}\cdot
                                            \E{\I{e_0[\harpoon\imath]}{e_1[\harpoon\imath]}}
                                                                        &(e_0,e_1:[\harpoon n]T)
\end{align*}
\end{framed}
\caption{Grammar and Semantics for the extension of the language with inner-products}
\label{fig:adjoint_grammar}
\endgroup
\end{figure}

To do this, we will define adjoint differentiation as rewriting expressions of the form
\begin{align*}
    \kwd{let} X_0 &\keq \hdots \kwd{in} \\
    \kwd{let} X_1 &\keq \hdots \kwd{in} \\
                  &\vdots \\
    \mwd{let} dX_0 &\meq \hdots \mwd{in} \\
    \mwd{let} dX_1 &\meq \hdots \mwd{in} \\
                  &\vdots \\
            & \inner{e_0}{e_1}_{\msub{p}} \mplus \cdots
\end{align*}
This form is defined as an extension of our language (Figure \ref{fig:adjoint_grammar}) with semantics given by de-sugaring the additional constructs into our original language.  However, at the beginning of this process, we will have a single inner-product with the beginning term on the right side; and at the end, we will be able to re-accumulate the result into a single term on the left side of a single inner-product.  Doing this will require application of certain structural rules. (rules \ref{rule:adj-unpack-let}--\ref{rule:adj-merge_pairs})

\begin{figure}[t]
\begingroup
\newcommand{\I}{\inner}
\newcommand{\D}{\deriv}
\newcommand{\+}{\mplus}
\newcommand{\0}{\mzero}
\newcommand{\h}{\harpoon}
\newcommand{\hi}{\harpoon\imath}
\newcommand{\Ip}[3]{\inner{#1}{#2}_{\!\msub{#3}}}
\arraycolsep=1.8pt
\begin{framed}
\begin{align}
\label{rule:adj-unpack-let}
\I{dY}{\D{\kwd{let} X \keq e_0 \kwd{in} e_1}{\sigma}}
                          &\rewritesto
  \begin{array}[t]{rrcl}
    \kwd{let}& X  &\keq& e_0 \kwd{in} \\
    \mwd{let}& dX &\meq& \D{e_0}{\sigma} \mwd{in} \\
    & \multicolumn{3}{l}{\I{dY}{\D{e_1}{\sigma\msub{X \mapsto dX}}}} \\
  \end{array} \\
\label{rule:adj-swap-let}
\begin{array}{rrcl}
  \mwd{let}& dX &\meq& \D{e_0}{\sigma} \mwd{in} \\
  \kwd{let}& X  &\keq& e_1 \kwd{in} e^L\\
\end{array}
  &\rewritesto
  \begin{array}{rrcl}
  \kwd{let}& X  &\keq& e_1 \kwd{in} \\
  \mwd{let}& dX &\meq& \D{e_0}{\sigma} \mwd{in} e^L \\
  \end{array} \\
\label{rule:adj-split-pairs}
\inner{dY}{\D{(e_0,e_1)}{\sigma}}
    &\rewritesto \inner{\pi_0 dY}{\D{e_0}{\sigma}} \+ \inner{\pi_1 dY}{\D{e_1}{\sigma}}\\
\label{rule:adj-proj-to-pair}
\inner{dY}{\D{\pi_0~e}{\sigma}}
    &\rewritesto \inner{(dY,0)}{\D{e}{\sigma}}\\
\label{rule:adj-merge_pairs}
\inner{(dY_0,0)}{e} \+ \inner{(0,dY_1)}{e}
    &\rewritesto \inner{(dY_0,dY_1)}{e}\\
\label{rule:adj-rev-let}
\begin{array}{l}
  \mwd{let} dX \meq \D{e_0}{\sigma} \mwd{in} \\
  \Ip{e_1}{dX}{p} \+ e^S
\end{array}
                          &\rewritesto
    \kwd{let} dX \keq e_1 \kwd{in} \Ip{dX}{\D{e_0}{\sigma}}{p} \+ e^S 
                                & dX \not\in FV(e^S) \\
\label{rule:adj-rev-let-dead-code}
\mwd{let} dX \meq \D{e_0}{\sigma} \mwd{in} e^S
                          &\rewritesto e^S 
                                & dX \not\in FV(e^S) \\
\label{rule:adj-diff-var}
\Ip{e}{\D{X}{\sigma[X\mapsto dX]}}{p}
                          &\rewritesto \Ip{e}{dX}{p} \\
\label{rule:adj-zero-var}
\Ip{e}{\D{X}{\sigma}}{p}  &\rewritesto \0 & X\not\in \sigma \\
\label{rule:adj-const}
\Ip{e}{\D{C}{\sigma}}{p}  &\rewritesto \0 \\
\label{rule:adj-plus-zero}
e^I \+ \0                 &\rewritesto e^I \\
\label{rule:adj-add-inner-prods}
\Ip{e_0}{dX}{p_0} \+ \Ip{e_1}{dX}{p_1}
                          &\rewritesto
    \begin{array}{l}
        \kwd{let} dY \keq \ind{p_0}\cdot e_0 + \ind{p_1}\cdot e_1 \kwd{in}\\
        \Ip{dY}{dX}{p_0\disj p_1}
    \end{array} \\
\label{rule:adj-add}
\Ip{dY}{\D{\begin{array}{l}
    \ind{p_0}\cdot X_0~ + \\
    \ind{p_1}\cdot X_1
\end{array}}{\sigma}}{p_2}
                          &\rewritesto
                \begin{array}{l} \Ip{dY}{\D{X_0}{\sigma}}{p_0\conj p_2} \+ \\
                                 \Ip{dY}{\D{X_1}{\sigma}}{p_1\conj p_2} \end{array}\\
\label{rule:adj-black-box-func}
\Ip{dY}{\D{\ind{p_0}\cdot f(X)}{\sigma}}{p_1}
                          &\rewritesto
    \begin{array}{l}
        \kwd{let} dX' \keq \ind{p_0\conj p_1}\cdot D^Tf(X, dY) \kwd{in}\\
        \Ip{dX'}{\D{X}{\sigma}}{p_0\conj p_1}
    \end{array}
\end{align}
\end{framed}
\caption{Adjoint Derivative of Tensors (first half)}
\label{fig:adjoint-deriv-1}
\endgroup
\end{figure}


\begin{figure}[t]
\begingroup
\newcommand{\I}{\inner}
\newcommand{\D}{\deriv}
\newcommand{\+}{\mplus}
\newcommand{\0}{\mzero}
\newcommand{\h}{\harpoon}
\newcommand{\hi}{\harpoon\imath}
\newcommand{\Ip}[3]{\inner{#1}{#2}_{\!\msub{#3}}}
\arraycolsep=1.8pt
\begin{framed}
\begin{align}
\label{rule:adj-binary-product}
\begin{split}
&\Ip{dY}{\D{\gen_{\h g}\ssum_{\h\imath_0 \h\imath_1 \h s}\ind{p}\cdot
                                X_0[\h\imath_0] \cdot X_1[\h\imath_1]}{\sigma}}{p'} \\
\rewritesto &
    \kwd{let} dX_0' \keq
        \gen_{\h\imath_0}\ssum_{\h g \h\imath_1 \h s}\ind{p\conj p'}\cdot
                                dY[\h g] \cdot X_1[\h\imath_1] \kwd{in} \\
  & \kwd{let} dX_0' \keq
        \gen_{\h\imath_1}\ssum_{\h\imath_0 \h g \h s}\ind{p\conj p'}\cdot
                                X_0[\h\imath_0] \cdot dY[\h g] \kwd{in} \\
  & \Ip{dX_0'}{\D{X_0}{\sigma}}{p_0} \+ \Ip{dX_1'}{\D{X_1}{\sigma}}{p_1} \\
    &\ \ \ \ \left(where \begin{array}{rcl}
                    p_0 &=& \exquant_{\h g \h\imath_1 \h s},~ p\conj p' \\
                    p_1 &=& \exquant_{\h\imath_0 \h g \h s},~ p\conj p'
             \end{array}\right)
\end{split}
\end{align}
\begin{align}
\label{rule:adj-unary-product}
\begin{split}
\Ip{dY}{\D{\gen_{\h g}\ssum_{\hi \h s}\ind{p}\cdot X[\hi]}{\sigma}}{p'} &\rewritesto
        \kwd{let} dX' \keq \gen_{\hi}\ssum_{\h g \h s}\ind{p\conj p'}\cdot dY[\hi] \kwd{in} \\
        &\phantom{\rewritesto}
        \Ip{dX'}{\D{X}{\sigma}}{\exquant_{\h g \h s},~p\conj p'}
\end{split}\end{align}
\end{framed}
\caption{Adjoint Derivative of Tensors (second half)}
\label{fig:adjoint-deriv-2}
\endgroup
\end{figure}

In order to get a handle on the rules for adjoint differentiation (Figures \ref{fig:adjoint-deriv-1} \& \ref{fig:adjoint-deriv-2}), it's worth detailing the specific way and order in which they are applied.  To begin with, rule \ref{rule:adj-unpack-let} is repeatedly applied until all the $\kwd{let}$-bindings have been extracted from the inner product and duplicated into $\mwd{let}$-bindings of differentials.  These can then be re-ordered with all differential bindings coming later by using rule \ref{rule:adj-swap-let}.  At this point, if the ``return value'' of the expression has a structured, tuple-type, then rule \ref{rule:adj-split-pairs} is applied, until the single inner-product of tuples is decomposed into a sum of inner-products, each of non-tuple values.  (There are now no tuples anywhere in the expression, other than the left-hand-side projections.)  At this point, the main body of the adjoint differentiation begins.

Before detailing the main activity of adjoint differentiation, let us skip ahead to the end, when all unprocessed $\mwd{let}$ bindings have been eliminated in favor of the final $\kwd{let}$ bindings.  At this point, there may be a sum of multiple inner products left if the original expression had more than one free (input) variable, or input variables of structured (pair) type.  In order to collect all these (now) outputs together, we must apply rules \ref{rule:adj-proj-to-pair} \& \ref{rule:adj-merge_pairs}.  After doing that, there will remain at most one inner product for each input/free-variable in the original expression.  The left-hand-sides of each of these inner products can then be joined together into a single left-hand-side output tuple, wrapped by all of the accumulated $\kwd{let}$ bindings.  That expression is the result.

The main body of adjoint differentiation works by applying the remaining rules, crucially rule \ref{rule:adj-rev-let}.  This is the reversing rule that produces the pattern characteristic of ``reverse-mode'' differentiation or ``back-propagation''.  It can be justified straight-forwardly as a $\beta$-substitution of $dX$ according to the semantics of $\kwd{let}$ bindings, followed by a reverse-$\beta$-substitution re-using the same variable name.  In the case of an un-used differential value (rule \ref{rule:adj-rev-let-dead-code}), we may simply drop the un-used $\mwd{let}$-binding instead, as an optimization.  As we go along, various rules in addition to multiple use of intermediates will produce larger sums of inner products.  Note that we are only allowed to apply rule \ref{rule:adj-rev-let} if there is exactly one inner product with a given right-hand-side differential variable.  This can be achieved by application of the rule \ref{rule:adj-add-inner-prods} to re-group multiple contributions.

With these structural issues cleared up, the remaining adjoint-differentiation rules reduce to handling the particular kinds of primitive forms allowed in our tensor SSA.  While these rules can get lengthy, their correctness can be verified by simple algebraic manipulation.  Likewise, the correctness of the overall scheme is verifiable transitively by verifying the correctness of each individual rule.

\begin{figure}[t]
\begingroup
\newcommand{\I}{\inner}
\newcommand{\+}{\mplus}
\newcommand{\0}{\mzero}
\newcommand{\E}{\CostExpr}
\newcommand{\D}{\SpanCost}
\arraycolsep=1.8pt
\begin{framed}
\begin{align}
\E{\mwd{let} dx \meq e \mwd{in} e^L}{\sigma} &= \E{e}{\sigma} + \E{e^L}{\sigma} \\
\E{e^I \+ e^S}{\sigma}          &= \E{e^I}{\sigma} + \E{e^S}{\sigma} \\
\E{\0}{\sigma}                  &= 0 \\
\begin{split}
\E{\I{e_0}{e_1}}{\sigma}        &= \$(T) + \E{e_0}{\sigma} + \E{e_1}{\sigma} \\
                                &\ \ \ (e_0,e_1 : T) \\
&\ \ \ \$(\R) = 1 \\
&\ \ \ \$(T_0,T_1) = \$(T_0) + \$(T_1) \\
&\ \ \ \$([n]T) = n \cdot \$(T)
\end{split} \\
\E{\I{e_0}{e_1}_{\!\msub{p}}}{\sigma}
                                &= \left(\sum_{\harpoon\imath}\ind{p}\right) + \E{e_0}{\sigma} + \E{e_1}{\sigma}
                                & (e_0,e_1 : [n_0,\hdots,n_k]\R) 
\end{align}
\end{framed}
\caption{Cost-Model Extension for Inner-product expressions}
\label{fig:adjoint_cost}
\endgroup
\end{figure}

In order to make a cost-preservation argument for reverse-mode automatic differentiation, we must somehow extend the cost model alongside the extended term grammar.  One natural way to do this would be to cost terms according to their ``desugared'' cost.  However, doing this will not produce the correct invariance for our argument to work out.  Instead, we must rely on a slightly different approach (Figure \ref{fig:adjoint_cost}).  In effect, we cost the overall term by summing the cost of all the constituent expressions, plus an adjustment for the sizes of all the current inner-products.  Crucially, observe that this adjustment for inner-products is not based on the cost of computing an inner product, but rather on the size of the non-zero interface between the left and right hand sides.

\begin{thm}[cost-efficiency of Adjoint Derivative]
Let $e$ be an expression in Tensor SSA form.  Let $T_I = (T_0,T_1,\hdots)$ be the ``input type'' of the free variables of $e$.  Let $T_O$ be the ``output type'' of $e$.  Let $\$(T)$ be the size-cost of a type as defined in Figure \ref{fig:adjoint_cost}.  Let $\$_{IO}(e) = \CostExpr{e}{} + \$(T_I) + \$(T_O)$ be the input-output-sensitive cost.  Then $\$(\derivT{e}{\sigma}) \leq 4\cdot \$(e)$.
\end{thm}
\begin{proof}
The proof proceeds by showing that the rewriting system of Figures \ref{fig:adjoint-deriv-1} \& \ref{fig:adjoint-deriv-2} preserves the extended cost model of Figure \ref{fig:adjoint_cost}.  This can be done by proving preservation for each individual rule (see following).  Then, (letting $x' : T_I$ stand in for the free variables of $e$ and $y' : T_O$) the final composition of these rewrites establishes the cost-equivalency bound $\inner{\derivT{e}{\sigma}}{x'} \leq \inner{y'}{\deriv{e}{\sigma}} + 2~\$(T_I)$.  To show our final desired bound, observe that
\begin{align*}
 \$_{IO}(\derivT{e}{\sigma})
    &= \CostExpr{\derivT{e}{\sigma}}{} + \$(T_I) + \$(T_O) \\
    &= \CostExpr{\inner{\derivT{e}{\sigma}}{x'}}{} + \$(T_O) \\
    &\leq \CostExpr{\inner{y'}{\deriv{e}{\sigma}}}{} + 3~\$(T_I) + \$(T_O) \\
    &= \CostExpr{\deriv{e}{\sigma}}{} + 3~\$(T_I) + 2~\$(T_O) \\
    &\leq \CostExpr{\deriv{e}{\sigma}}{} + 4(\$(T_I) + \$(T_O)) \\
    &\leq 4~\CostExpr{e}{} + 4(\$(T_I) + \$(T_O)) \\
    &= 4\cdot\$_{IO}(e)
\end{align*}
where the last inequality follows from Theorem \ref{thm:fwd-deriv-cost}.

We have provided the overall structure of the proof.  All that remains is to verify that each individual rule does in fact preserve cost.  In general, we will usually have to apply the total derivative to the left-hand-side of a rule in order to expose the comparable terms to cost.  For instance, rule \ref{rule:adj-unpack-let} is trivial to verify after we apply this procedure.  Rule \ref{rule:adj-swap-let} is even more trivial to check.  Rule \ref{rule:adj-split-pairs} follows from the cost structure of inner-products and pair types.  Rules \ref{rule:adj-proj-to-pair} \& \ref{rule:adj-merge_pairs} are applied successively, and only at the very end of rewriting.  While rule \ref{rule:adj-proj-to-pair} increases cost, these costs are then subsequently eliminated by rule \ref{rule:adj-merge_pairs} when contributions to both sides of a pair can be found.  In the odd cases where they cannot be found, the additional cost attributed to taking inner products with zeros can be absorbed in the $\cdots + \$(T_I)$ cost-allowance we provided earlier.  Likewise, we can drop the inner-product filtering predicate $p$ at this final output stage, which may result in up to another additional $\$(T_I)$ cost-adjustment.  These are the only two such cost-adjustments we must make.

Rule \ref{rule:adj-rev-let} simply re-arranges terms from a cost perspective.  Rule \ref{rule:adj-rev-let-dead-code} only eliminates sub-terms, and so may only decrease cost.  Rules \ref{rule:adj-diff-var}--\ref{rule:adj-plus-zero} are all trivially cost-neutral.

The remaining rules specify behavior for the costly SSA operations rather than simply different kinds of structural book-keeping.  These rules will rely on shifting cost between the expressions and the inner-products to maintain a careful balance that relies on some of the more subtle details of our formal cost model.

To show cost-preservation for rule \ref{rule:adj-add-inner-prods}, we recall that the addition operation only counts towards the cost when $p_0 \conj p_1$ is true, and then apply the inclusion-exclusion principle.  To be more detailed, first observe the invariant in our rewrite rules that the left-hand-side of any inner product is always an expression of cost $0$.  Therefore the cost of $e_0$ and $e_1$ are both $0$.  The cost prior to a rewrite is therefore $\#(p_0) + \#(p_1)$, where $\#(p) = \sum_{\harpoon\imath} [p]$.  On the right-hand-side, the inner product now has cost $\#(p_0\disj p_1)$, while the let-binding has cost $\#(p_0\conj p_1)$ according to the aforementioned cost rule.  By the inclusion-exclusion principle, these are equal.

To show cost-preservation for rule \ref{rule:adj-add}, we use similar reasoning.  On the left here, we have a cost of $\#(p_2)$ due to the inner product and $\#(p_0\conj p_1)$ due to the guarded addition.  On the right, we have the cost $\#(p_0\conj p_2) + \#(p_1\conj p_2) = \#(p_0\conj p_1 \conj p_2) + \#((p_0\disj p_1)\conj p_2)$, by inclusion-exclusion.  The two terms after applying inclusion-exclusion are bounded $\#(p_0\conj p_1) \geq \#(p_0\conj p_1 \conj p_2)$, and $\#(p_2) \geq \#((p_0\disj p_1)\conj p_2)$, so that the cost after rewriting is no more than the cost before.

Rule \ref{rule:adj-black-box-func} follows a similar argument, except the bound is even more straight-forward. (no need for inclusion-exclusion)

\begingroup
\newcommand{\hg}{\harpoon g}
\newcommand{\hi}{\harpoon\imath}
\newcommand{\hs}{\harpoon s}

It is easier to examine rule \ref{rule:adj-unary-product} before rule \ref{rule:adj-binary-product}, since it already requires all of the important ideas in the more complicated binary case.  To start, look at the cost on the right-hand-side, which consists of $\sum_{\hi}\left[ \exists_{\hg\hs} p\conj p' \right]$ for the inner-product and $\sum_{\hi\hg\hs}[p\conj p'] - \sum_{\hi}\left[ \exists_{\hg\hs} p \conj p' \right]$ for the let-binding.  Adding these two together, we see that the cost adjustment for big-summations and the cost of the inner product perfectly cancel.  So we are left with a right-hand-side cost of $\$RHS = \sum_{\hi\hg\hs}[p\conj p']$.
On the left-hand-side observe that the cost is $\$LHS = \sum_{\hg} [p'] + \sum_{\hg\hi\hs} [p] - \sum_{\hg}\left[ \exists_{\hi\hs} p \right]$.  If there were no inner-product or big-summation adjustment terms, then we would be done, since $[p] \geq [p\conj p']$.  However, the subtraction of the adjustment raises the possibility that somehow the cost on the left is less than the cost on the right.  To see why this can't be the case, observe that
\begin{align*}
\$LHS
&= \sum_{\hg\hi\hs} [p] + \sum_{\hg} [p'] - \sum_{\hg}\left[ \exists_{\hi\hs} p \right] \\
&\geq \sum_{\hg\hi\hs} [p] - \sum_{\hg}\left[ \exists_{\hi\hs} p \conj \neg p' \right] \\
&\geq \sum_{\hg\hi\hs} [p] - \sum_{\hg\hi\hs} [p\conj \neg p'] \\
&= \sum_{\hg\hi\hs}\left( [p] - [p\conj \neg p'] \right) \\
&= \sum_{\hg\hi\hs}[p\conj p']\\
&= \$RHS
\end{align*}

\endgroup

Rule \ref{rule:adj-binary-product} follows according to the same argument, with additional accounting for multiplications.  We have now shown cost-equivalency for all of the rules.
\end{proof}

\clearpage

\begin{acks}
The authors would like to thank Zach Devito, Michael Bakkemo, Matt Kehrt for conversations about this project and automatic differentiation more generally.  We would also like to thank Alex Reinking for some LaTeX tricks.
\end{acks}

\bibliographystyle{ACM-Reference-Format}
\bibliography{ref}


\end{document}